\documentclass[12pt]{article}
\usepackage{psfrag,epsfig,amsmath,amssymb}

\makeatletter
\renewcommand{\section}{\setcounter{equation}{0}\@startsection
  {section}%
  {1}%
  {0pt}%
  {-1\baselineskip}%
  {0.4\baselineskip}%
  {\bfseries\large}}%
\renewcommand{\subsection}{\@startsection
  {subsection}%
  {2}%
  {0pt}%
  {-0.75\baselineskip}%
  {0.2\baselineskip}%
  {\bfseries}}%
\renewcommand{\subsubsection}{\@startsection
  {subsubsection}%
  {3}%
  {0pt}%
  {-0.5\baselineskip}%
  {0.1\baselineskip}%
  {\sc}}%
\makeatother

\makeatletter
 \newcommand\figcaption{\def\@captype{figure}\caption}
\makeatother

\def\a{\alpha}
\def\b{\beta}
\def\d{\delta}

\def\ga{\gamma}

\def\la{\lambda}

\def\m{\mu}
\def\n{\nu}
\def\r{\rho}
\def\s{\sigma}

\def\th{\theta}

\def\Dirac{{D\mkern-12mu/}}

\def\prslash{{\partial\mkern-9mu/}}
\def\pslash{{p\mkern-8mu/}{\!}}

\def\prslash{{\partial\mkern-9mu/}}    

\def\pslash  {{p\mkern-7mu/}}

\def\bp{\text{\tiny{BPST}}}
\def\id{{\rm{I}\!\rm{I}}}

\def\id3x{\int\!\! d^3\!\vec{x}}
\def\idx{\int\!\! d^4\!x}
\def\iDx{\int\!\! d^D\!x}

\def\rig>{\right>}

\newcommand{\bea}{\begin{eqnarray}}
\newcommand{\eea}{\end{eqnarray}}
\newcommand{\beann}{\begin{eqnarray*}}
\newcommand{\eeann}{\end{eqnarray*}}
\newcommand{\ba}{\begin{array}}
\newcommand{\ea}{\end{array}}

\newcommand{\Tr}{\mathbf{Tr}}

\def\Psib{\bar{\Psi}}
\def\g5{\gamma_{5}}

\def\prslash {{\partial\mkern-9mu/}}  
\def\pslash  {{p\mkern-7mu/}}

\def\idx3{\int\! d^{3}\!\vec{x}\,}
\def\idx{\int\! d^{4}\!x\,}

 \def\Psib{\bar{\Psi}}

 \def\Dirac{{D\mkern-12mu/}\,}

 \def\prslash {{\partial\mkern-9mu/}}  
 \def\pslash  {{p\mkern-7mu/}}

\def\bk {\bar{k}} 
 \def\bp {\bar{p}}
 \def\bq {\bar{q}} 
 \def\bg {\bar{\gamma}}
 \def\bdelta {\bar{\delta}}

 \def\Db {{\partial}_{\beta}}
 \def\Dm {{\partial}_{\mu}}
 
 \def\ab {a_{\beta}}
 \def\am {a_{\mu}}

 \def\g {\gamma}

 \def\mi {{\mu_1}}

 \def\a {\alpha}
\def\b {\beta}
\def\r {\rho}
 \def\s {\sigma}

 \def\Tr{\text{Tr}}

\begin{document}
\begin{titlepage}
\hfill{NSF-KITP-09-188}\\
\rightline{FTI/UCM 107-2009}\vglue 10pt
\begin{center}

{\Large \bf Noncommutative  GUT inspired theories with U(1), SU(N) groups and their renormalisability}\\
\vskip 0.2 true cm 
\vfill
 C. Tamarit\footnote{E-mail: tamarit@kitp.ucsb.edu}
\vskip 1pt  
 {\it Kavli Institute for Theoretical Physics, University of California\\
 Santa Barbara, CA, 93106-4030, USA}\\
\end{center}
\vfill
{\leftskip=50pt \rightskip=50pt \noindent We consider the GUT compatible formulation of noncommutative QED, as well as noncommutative SU(N) GUTs, for $\rm N>2$, with no scalars but with fermionic matter in an arbitrary, anomaly-free representation, in the enveloping algebra approach. We compute, to first order in the noncommutativity parameters $\th^{\m\n}$, the UV divergent part of the one-loop background-field effective action involving at most two fermion fields and an arbitrary number of gauge fields. It turns out that, for special choices of the ambiguous trace over the gauge degrees of freedom, for which the $O(\th)$ triple gauge-field interactions vanish, the divergences can be absorbed by means of multiplicative renormalisations and the inclusion of $\th$-dependent counterterms that vanish on-shell and are thus unphysical. For this to happen in the SU(N), ${\rm N}>2$ case, the representations of the matter fields must have a common second Casimir;  anomaly cancellation then requires the ordinary (commutative) matter content to be non-chiral. Together with the vanishing of the divergences of fermionic four point functions, this shows that GUT inspired theories with U(1) and SU(N), $\rm N>2$ gauge groups and ordinary vector matter content not only have  a renormalisable matter sector, but are on-shell one-loop multiplicatively renormalisable at order one in $\theta$.

\par }

\vfill
\end{titlepage}


\setcounter{page}{2}
\section{Introduction}

     The enveloping algebra approach, which makes use of Seiberg-Witten maps, is the only known approach that allows to construct noncommutative  gauge theories with arbitrary groups and representations  \cite{Jurco:2001rq}. Much work has been carried out to analyse the properties of these theories, in particular pertaining to their consistency at the quantum level: though the models involve an infinite number of interactions that are not power-counting renormalisable, their properties at the quantum level are better than naively expected. This includes anomaly cancellation conditions, which have been shown to be identical to those in commutative gauge theories \cite{Brandt:2003fx}, and also renormalisability properties. In this respect, the more striking result could be the observed renormalisability of the gauge sector at one-loop, for several models with diverse matter content \cite{Wulkenhaar:2001sq,Buric:2002gm,Buric:2004ms,Buric:2005xe,Latas:2007eu,Buric:2006wm,Martin:2006gw,Martin:2009mu}; in fact, the matter determinants contributing to the one-loop gauge effective action are known to yield renormalisable contributions to all orders in $\theta$ \cite{Martin:2007wv}. 
     
     Regarding the renormalisability of theories with fermionic matter, generically the inclusion of noncommutative Dirac fermion fields gives rise to four fermion UV divergences that spoil the renormalisability; the lack thereof has been shown explicitly for the non-GUT-compatible version of QED of ref.~\cite{Wulkenhaar:2001sq} and for the SU(2) gauge theory with fundamental  fermions \cite{Buric:2004ms}, but the problematic four fermion divergences have been shown to appear in generic theories with noncommutative Dirac fermions in arbitrary representations \cite{Martin:2009sg}. However, GUT-compatible theories, which were introduced in ref.~\cite{Aschieri:2002mc}, have been shown to be free of these divergences  for arbitrary choices of the representation of the fermion fields  \cite{Martin:2009sg,Buric:2007ix}. Moreover,  anomaly safe GUTs have been shown  to have, against all odds, a one-loop renormalisable  effective action at order one in the noncommutativity parameters $\theta$  \cite{Martin:2009vg}; these are the first noncommutative gauge theories with fermionic matter in representations other then the adjoint that have been shown to be one-loop renormalisable. The only other known examples of one-loop, $O(\th)$ renormalisable noncommutative gauge theories defined by means of Seiberg-Witten maps and involving fermion fields are (S)U(N) SuperYang Mills theories \cite{Martin:2009mu}.
     
     In this paper we continue the study of the renormalisability properties of noncommutative GUT compatible theories with no scalar fields by analysing models in which the gauge group is not anomaly safe, i.e., the anomaly coefficients are non-vanishing, so that anomaly cancellation conditions restrict the allowed representations, which must be reducible for the total contribution to the anomaly to vanish. The main difference with the theories analysed in ref.~\cite{Martin:2009vg} is, apart from the restrictions in the matter content coming from the gauge anomaly, that  local gauge invariant contributions of the form $\Tr\theta fff$, with $f$ being field strengths, are allowed in both the tree-level and effective actions. In fact, the tree-level action does include these $O(\th)$ bosonic interactions, which are sensitive to ambiguities in the definition of the trace over the gauge degrees of freedom; this has the effect of introducing an additional coupling. We consider, in particular, the GUT-inspired version of QED \cite{Aschieri:2002mc}, and SU(N) GUT theories, for $\rm N>2$, with fermions in a generic, anomaly free representation; it should be recalled that the non-GUT-compatible counterparts of these theories, formulated in terms of noncommutative Dirac fermions, are not renormalisable. To analyse the renormalisability of the matter sector of these theories, completing the results concerning four fermion UV divergences of ref.~\cite{Martin:2009sg}, we compute the UV divergent part of the one-loop effective action at order $\theta$ involving two fermion fields. In order to address the full one-loop renormalisability, we also compute the UV  divergences in the bosonic sector. We use  the background field method in dimensional regularisation, which  allows to reconstruct the full one-loop, $O(\theta)$, non-evanescent divergent contributions involving two or no fermion fields from the computation of the pole part of the two- and three-point Green functions involving, respectively, two fermion fields, one gauge field and two fermion fields, and three gauge fields. 
     
     The results are the following: both in the QED and SU(N), $\rm N>2$ cases, the divergences can be subtracted by means of multiplicative renormalisations and the inclusion of redundant interactions in the form of $\th$-dependent counterterms that vanish on-shell, whenever the ambiguous trace over the bosonic degrees of freedom is chosen so that the $O(\theta)$ bosonic interactions cancel --making the situation similar to that of anomaly safe GUTs ~\cite{Martin:2009vg}. In the SU(N) case, there is a further requirement for renormalisability, which is that all the irreducible matter representations must share the same second Casimir; anomaly cancellation requires then the ordinary matter content to be nonchiral, consisting of two multiplets in representations that are conjugate of each other --note, however, that the noncommutative interactions remain nonchiral in the sense that they cannot be written in terms of noncommutative Dirac fermions. Together with the results concerning the absence of four fermion UV divergences in the effective action, this shows that the models with ordinary nonchiral matter content not only have a renormalisable matter sector, but the full one-loop, $O(\th)$ effective action is renormalisable. This renormalisability holds for the off-shell effective action, but in such a way that no additional physical parameters have to be introduced beyond those already present at tree level, which are the gauge couplings and the noncommutativity parameters. This is due to the fact that the counterterms come only from multiplicative renormalisations of the tree-level fields and parameters and from redundant interactions that vanish on-shell and hence depend on unphysical couplings. The on-shell effective action of the nonchiral theories, from which the S-Matrix elements derive, can thus be renormalised simply by means of multiplicative renormalisations of the fields and the tree-level physical couplings. The one-loop renormalisation does not require the addition of more physical parameters; this aspect is key in raising hopes that these theories might be truly renormalisable and predictive when higher order corrections are included, yielding finite S-Matrix elements depending on a finite number of physical couplings.

     These results allow to overcome the nonrenormalisability observed for both noncommutative QED and SU(N), $\rm N>2$ gauge theories formulated in terms of noncommutative Dirac fermions --and thus not GUT-compatible-- with a nonchiral ordinary matter content. This adds more models to the list of one-loop, $O(\th)$ renormalisable noncommutative gauge theories with fermion fields defined by means of Seiberg-Witten maps. This list, empty until very recently, includes (S)U(N) super Yang-Mills theories, \cite{Martin:2009mu}, anomaly-safe GUTs with matter in an arbitrary irreducible representation \cite{Martin:2009vg}--with the possibility of adding fields in the conjugate representation-- and  the models found in this paper.

	The organisation of the paper is as follows.  The U(1) and SU(N), $N>2$ theories are defined in section 2, which also outlines the computation by means of the background field method. Section 3 includes the results of the computations of the divergent part of the effective actions involving two fermion fields, whereas section 4 deals with the divergences in the bosonic sectors. Renormalisability is analysed in section 5, and the results are discussed in section 6. Two appendices are also included: appendix A shows the Feynman rules relevant to the calculations, and appendix B displays the $\beta$ functions of the physical couplings of the theories.
	

	\section{The models and the background field method}
	
	We consider  four-dimensional noncommutative GUT-compatible theories with gauge groups U(1) and SU(N), $\rm N>2$,   and fermionic 
matter.  Paralleling the discussion in ref.~\cite{Martin:2009vg}, these theories are defined by means of a noncommutative left-handed chiral multiplet $\Psi$ in an arbitrary representation $\rho_\Psi$ of the gauge group, and an enveloping-algebra valued gauge field $A_\m$ with action~\cite{Aschieri:2002mc}
\begin{align}\label{S}
&{S}=\idx-\frac{1}{2\la^2}\Tr F_{\m\n}\star F^{\m\n}+\Psib_L i\Dirac \Psi_L,\\
&\nonumber F_{\m\n}=\partial_\m A_\n-\partial_\n A_\m-i[A_\m, A_\n]_\star,\quad D_{\m}\psi_L=\partial_\m\Psi_L-i \rho_{\Psi}(A_\m)\star\Psi_L.
\end{align}
The noncommutative fields are given in terms of the ordinary ones $a_\m,\psi$ by the Seiberg-Witten maps that follow,
\begin{align}
\nonumber A_\mu&=a_\m+\frac{1}{4}\th^{\a\b}\{\partial_\a \am+f_{\a\m},\ab\}+O(\th^2),\\
\label{SW}
\Psi_L&=\psi_L-\frac{1}{2}\th^{\a\b}\rho_\psi(a_\a)\Db\psi_L+\frac{i}{4}\th^{\a\b}\rho_\psi(a_\a) \rho_\psi(a_\b)\psi_L+O(\th^2).
\end{align}
$\rho_{\psi}$  might be expressed as a direct sum of irreducible representations, $\rho_{\psi}=\bigoplus_{r=1}^{N_F} \rho_{r}$.  Accordingly, the fermion fields can be expressed as a direct sum of irreducible multiplets, $\Psi_{L}=\bigoplus_{r=1}^{N_F} \Psi_{L}^{r}$, $\psi_{L}=\bigoplus_{r=1}^{N_F} \psi_{L}^{r}$. The ordinary field $a_\m$ takes values in the Lie algebra; we will use the following notations for the expansion in terms of generators,
\begin{align}\label{genexp}
{\rm U(1):  }&\quad a_\m=e a_\m^0 Y,\\
\nonumber{\rm SU(N):}& \quad a_\m=a_\m^a T^a, \,\,a=1,\dots,N^2-1.
\end{align}
We furthermore consider the following normalisation for the generators in an arbitrary representation: 
\begin{align}\label{rengen}
\rho_r(Y)^2\equiv Y_r^2=1,\quad \Tr\{\rho_r(T^a),\rho_r(T^b)\}\equiv \Tr_r\{T^a,T^b\}=c(r)\delta^{ab},
\end{align}
$c(r)$ denoting the Dynkin index of the representation $r$ of SU(N). The matter content is chosen so that it is anomaly free. In the U(1) case, to make contact with QED, we consider a multiplet with two irreducible representations $\Psi^\pm$, with $\rho_+(Y)\equiv Y_+=1$ and $\rho_-(Y)\equiv Y_-=-1$.  In the SU(N) case the representation is kept arbitrary, though with a vanishing total anomaly.

 The action in eq.~\eqref{S} has an ambiguity which should not be ignored, and which affects the trace in the noncommutative gauge  kinetic term: since the noncommutative fields are enveloping-algebra valued, the result of the trace is representation dependent. This ambiguity was irrelevant in the case of GUT theories with anomaly safe groups, since the $O(\theta)$ corrections to the tree-level bosonic action are proportional to $\Tr\{T^a,T^b\}T^c$, which vanishes for those theories but not for the groups U(1), SU(N), $\rm N>2$ considered in this paper \cite{Aschieri:2002mc}. We define $\Tr$ in the gauge kinetic term as a sum of the traces along the different representations,
\begin{align}\label{sumtraces}
\Tr=\sum_r \kappa_r \Tr_r.
\end{align}
$\Tr_r$ is defined in the nonabelian case in eq.~\eqref{rengen}. In the U(1) case, $\Tr_r$  is understood as the application of $\rho_r$.
It is natural to consider in the sum of eq.~\eqref{sumtraces} exclusively the representations of the matter fields;  the inclusion of these --and no others-- can be justified from demanding the renormalisability of the matter determinants that contribute to the bosonic effective action \cite{Martin:2007wv}.  Note that the inclusion of a finite number of representations in the sum is equivalent to consider a finite number of couplings in the gauge sector, which include the ordinary gauge coupling present in commutative theories. More precisely, to make contact with the commutative limit, and taking into account the normalisation in eq.~\eqref{rengen}, we demand
 \begin{align}\label{g}
 {\rm U(1)}: \frac{\sum_\pm \kappa_\pm (e Y_\pm)^2}{\la^2}=\frac{e^2(\kappa_++\kappa_-)}{\la^2}=\frac{1}{2},\quad {\rm SU(N)}:
\frac{\sum_r \kappa_r c(r)}{\lambda^2}=\frac{1}{2g^2},
\end{align}
where $g$ is the commutative SU(N) coupling.

The $O(\th)$ pure gauge contribution to the action is then, using eqs.~\eqref{S}, \eqref{SW} and \eqref{g}, of the form
\begin{align}\label{Sncbos}
S^{\rm NC}_{\rm bos}&=\idx \frac{\sum_r \kappa_r }{\la^2}\Tr_r\Big(\frac{1}{4}\th^{\a\b}f_{\a\b}f_{\m\n}f^{\m\n}-\th^{\a\b}f_{\m\a}f_{\n\b}f^{\m\n}\Big)+O(h^2),\\
\nonumber &f_{\m\n}=\partial_\m a_\n-\partial_\n a_\m-i[a_\m,a_\n].
\end{align}
In the U(1) case, using the notation in eq.~\eqref{genexp}, we define
\begin{align}\label{f0}
 f_{\m\n}^{\rm U(1)}=e Y (f_{\m\n})^0,\,(f_{\m\n})^0=\partial_\m a^0_\n-\partial_\n a^0_\m;
\end{align}
then the noncommutative bosonic action turns out to be
\begin{align}\label{SncbosU1}
S^{\rm NC,U(1)}_{\rm bos}=\idx \frac{\sum_\pm \kappa_\pm (eY_\pm)^3}{\la^2}\Big(\frac{1}{4}\th^{\a\b}(f_{\a\b})^0(f_{\m\n})^0(f^{\m\n})^0-\th^{\a\b}(f_{\m\a})^0(f_{\n\b})^0(f^{\m\n})^0\Big)+O(h^2).
\end{align}
In the SU(N) case, using 
\begin{align*}
\Tr_r T^a\{T^b,T^c\}=\frac{1}{2}A(r) d^{abc},
\end{align*}
where $A(r)$ is the anomaly coefficient of the representation $r$ (its value in the fundamental representation being $A(F)=1$), we can write the action \eqref{Sncbos} as
\begin{align}\label{SncbosSUN}
S^{\rm NC,SU(N)}_{\rm bos}&=\idx \frac{\sum_r \kappa_r A(r)}{\la^2}d^{abc}\Big(\frac{1}{16}\th^{\a\b}(f_{\a\b})^a (f_{\m\n})^b(f^{\m\n})^c\!-\!\frac{1}{4}\th^{\a\b}(f_{\m\a})^a(f_{\n\b})^b(f^{\m\n})^b\Big)+O(h^2).
\end{align}
We can regard the combinations $\frac{\sum_r \kappa_r (eY_r)^3}{\la^2}$  and $\frac{\sum_r \kappa_r A(r)}{\la^2}$ appearing in eqs.~\eqref{SncbosU1} and \eqref{SncbosSUN} as new couplings, since, as the sum over representations of the matter fields includes at least two different representations --a fact required by the anomaly cancellation conditions-- then the said combinations are independent of those appearing in eq.~\eqref{g}; we will denote these new couplings as
\begin{align}\label{gtilde}
\frac{\sum_\pm \kappa_\pm(eY_\pm)^3}{\la^2}=\frac{e^3(\kappa_+-\kappa_-)}{\la^2}\equiv\frac{\tilde e}{2},\quad \frac{\sum_r \kappa_r A(r)}{\la^2}\equiv \frac{1}{\tilde g^2}.
 \end{align}
 The definitions have been chosen in such a way that, when only the fundamental representation is chosen, one gets from eqs.~\eqref{g} and \eqref{gtilde} $\tilde e=e,\,\tilde g=g.$ The bosonic  
 actions end up being
\begin{align}
\nonumber S_{\rm bos}^{\rm U(1)}=&\idx -\frac{1}{4}g_{\m\n}g^{\m\n}+\tilde e\Big(\frac{1}{8}\th^{\a\b}(f_{\a\b})^0(f_{\m\n})^0(f^{\m\n})^0-\frac{1}{2}\th^{\a\b}(f_{\m\a})^0(f_{\n\b})^0(f^{\m\n})^0\Big)+O(h^2),\\
\nonumber S_{\rm bos}^{\rm SU(N)}\!=&\idx -\frac{1}{4g^2}(f_{\m\n})^a(f^{\m\n})^a+\frac{1}{\tilde g^2}d^{abc}\Big(\frac{1}{16}\th^{\a\b}(f_{\a\b})^a (f_{\m\n})^b(f^{\m\n})^c\!-\!\frac{1}{4}\th^{\a\b}(f_{\m\a})^a(f_{\n\b})^b(f^{\m\n})^c\Big)\\
&+O(h^2).\label{Sbos}
\end{align}
Note that, by choosing $\kappa+=\kappa_-$ and $\kappa_r=\kappa_{r^*}$, since the anomaly coefficients are such that $A(r)=-A(r^\star)$, one obtains $\tilde e=\frac{1}{\tilde g^2}=0$. Nevertheless, we will keep the couplings arbitrary, to make contact with previous studies in the literature in which they were taken as nonzero, and to see if renormalisability can be achieved for arbitrary values of $\tilde e, \tilde g$. On the other hand, in principle their cancellation needs not survive the renormalisation procedure, though in the end it will turn out that it does.

The expanded fermionic action is given for both theories by
\begin{align}\nonumber
S^{\rm NC}_{\rm fer}&=\sum_r  \idx\Big(\!-\frac{i}{4}\th^{\a\b}\bar\psi_r \rho_r(f_{\a\b}\Dirac)\,\psi_r-\frac{i}{2}\th^{\a\b}\bar\psi_r\ga^\m \rho_r(f_{\m\a}D_\b)\psi_r\Big)+O(\th^2),\\
D_\m\psi_r&=\Dm\psi_r-i\rho_r (\psi)_r.\label{Sncfer}
\end{align}

	We quantise the theory by means of path integral methods, defining the functional generator in terms of Feynman diagrams. In order to formulate the Feynman rules in terms of Dirac fermions, we add a spectator right-handed fermion, as in ref.~\cite{Martin:2009sg}
	\begin{equation*}
S\rightarrow S'= S+\idx\bar{\tilde\psi}_R i\prslash\tilde\psi_R,	\quad \psi=\left[\begin{array}{l}\tilde\psi_R\\
\psi_L\end{array}\right].
\end{equation*}
We regularise the theory, as in refs.~\cite{Martin:2009sg,Martin:2009vg} by means of dimensional regularisation in $D=4+2\epsilon$ dimensions,  using the BMHV scheme for defining $\gamma_5$  \cite{'tHooft:1972fi,Breitenlohner:1977hr}. The dimensionally regularised action is not unique in this scheme: there is an infinite number of actions which reduce to \eqref{S} in the limit $D=4$, differing by evanescent operators \cite{Martin:1999cc}. As in ref.~\cite{Martin:1999cc} we will treat the interaction vertices as ``four-dimensional'', meaning that we will keep all the vector indices in them contracted with the ``barred'' metric $\bar{g}_{\mu\nu}$. We will also define the dimensionally regularised $\theta^{\mu\nu}$ as ``four-dimensional''.  Furthermore, in our computations we will discard any contributions that have a pole in $\epsilon$ whose residue is an evanescent operator. These contributions involving evanescent operators have no physical effects at the one-loop level for an anomaly-free theory, and are therefore mere artifacts of the renormalisation procedure \cite{Martin:1999cc,Chanowitz:1979zu} --nevertheless, they should be taken into account in the computations at higher loop orders \cite{Martin:1993yx} .

We wish to obtain the divergent part of the effective action involving no evanescent operators, at most two fermion fields and an arbitrary number of gauge fields, in a manifestly covariant approach, which allows to reconstruct the full contribution to the effective action from a minimum number of diagrams, as was done in refs~\cite{Martin:2009mu,Martin:2009vg}.  For this we use the background field method  \cite{Abbott:1980hw}. Within this framework, the  the gauge field $a_\m$ is split in a background part $b_\m$ and a quantum part $q_\m,$
\begin{align*}
a_\mu=b_\m+q_\m,
\end{align*}
and a gauge fixing which preserves background gauge transformations is chosen  
$$\delta q_\m=-i[q_\m,c],\,\delta b_\m=D[b]_\m c,\,\,D[b]_\m=\partial_\m-i[b_\m,\,].$$
This gauge fixing is 
\begin{align*}
S_{gf}=-\frac{1}{2\alpha}\idx(D^{[b]}_\m q^\m)^2,\quad S_{gh}=\idx \bar cD^{[b]}_\m D^{[b+q]\m} c.
\end{align*}	
As discussed in ref.~\cite{Abbott:1980hw} --see also  \cite{Martin:2009mu}-- introducing the classical fields $\hat b_\m,\hat\psi$, the 1PI functional is given by
\begin{align}
\Gamma[\hat b_\m,\hat \psi,\hat{\bar\psi}]=\idx\sum_k\sum_n \frac{-i}{ (k!)^2}\tilde\Gamma^{(n,k)}_{\scriptsize\begin{array}{l}
i_1,..,i_k; \,\, j_1,..,j_k;\begin{array}{l}\mi,..,\mu_n\\
a_1,..,a_k\end{array}
\end{array}}\prod_{l=1}^k\hat{\bar\psi}_{i_l}\prod_{p=1}^k\hat\psi_{j_p}\prod_{m=1}^n\hat b_{\m_m}^{a_m} .
\label{Gammaexp}
\end{align}
The effective action above is gauge invariant under gauge transformations of the classical fields $\hat b_\m,\hat\psi,\hat{\bar\psi}$. Its dimensionally regularised version is only gauge invariant modulo an evanescent operator which, for the reasons stated before, can be ignored in one-loop calculations of UV divergences in an anomaly-free theory. $\tilde\Gamma^{(n,k)}$ is equivalent to a background 1PI diagram with $n$ background gauge field legs,  $k$ fermionic legs and $k$ anti-fermionic legs. (Note that our definitions do not involve any symmetrisation over the background gauge fields). The vertices relevant to our calculations and their associated Feynman rules for $\alpha=1$  are given in appendix A.
	 
As was done in refs.~\cite{Martin:2009mu,Martin:2009vg}, the computation is simplified by, on the one hand, choosing the  gauge $\alpha=1$,  and, on the other, by only calculating a minimum number of diagrams. The choice of $\alpha=1$ will not affect the on-shell effective action, which is independent of $\alpha$, and therefore the conclusions reached on-shell will have general validity \cite{Fradkin:1983nw}. The method of computing a minimum number of diagrams consists in reconstructing full gauge invariant contributions to the divergent part of the effective action by only calculating 1PI Green functions with the minimum number of fields appearing in those contributions. Since in the computation we ignore evanescent contributions that might break gauge invariance, and we are considering an anomaly-free matter content, the divergences we aim to obtain will be local gauge invariant polynomials in the field strength, fermion fields and their covariant derivatives. These UV divergent contributions can be expanded in a basis of independent gauge invariant terms; if the contributions to these terms with a given number and types of fields are also independent, then one can fix the coefficients in the expansion by computing the 1PI Green functions with the same number and types of fields.

To identify the diagrams that must be computed, as well as for the purpose of matching the counterterms needed to remove the divergences, we will use a basis of local gauge invariant terms whose integrals are independent. At first order in $\theta$, we might have purely bosonic terms, terms with two fermion fields and terms with four fermion fields.

For the bosonic terms we consider the following basis,
\begin{align}\label{rs}
r_1=&\th^{\a\b}\Tr_F\, f_{\a\b}\{f_{\m\n},f^{\m\n}\},\\
\nonumber r_2=&\th^{\a\b}\Tr_F\,f^{\m\n}\{f_{\m\a},f_{\n\b}\},
\end{align}
where $\Tr_F$ denotes the trace over the fundamental; in the U(1) case, we define $Y_F=+1$. In the case of terms with two fermion fields, we choose the following basis, for each flavour $r$:
\begin{align}
\nonumber s^r_1&=\th^{\a\b}\bar\psi_r\ga^\m P_L f_{\m\b}D_\a\psi_r,& s^r_2&=\th^{\a\b}\bar\psi_r\ga^\m P_L f_{\a\b}D_\m\psi_r, & s^r_3&=\th^{\a\b}\bar\psi_r\ga^\m P_L {\cal D}_\m f_{\a\b}\psi_r,\\
\nonumber s^r_4&=\th^{\a\b}\bar\psi_r\ga_\a P_L f_{\b\m}D^\m\psi_r,& s^r_5&=\th^{\a\b}\bar\psi_r\ga_\a P_L {\cal D}^\m f_{\b\m}\psi_r, &
s^r_6&=\th^{\a\b}\bar\psi_r{\ga_{\a\b}}^\m P_L {\cal D}^\n f_{\m\n}\psi_r,\\
\nonumber s^r_7&=\th^{\a\b}\bar\psi_r{\ga_{\a\b}}^\m P_L  f_{\m\n}D^\n\psi_r, & s^r_8&=\th^{\a\b}\bar\psi_r{\ga_{\a}}^{\r\s} P_L {\cal D}_\b f_{\r\s}\psi_r, & s^r_9&=\th^{\a\b}\bar\psi_r{\ga_{\a}}^{\r\s} P_L  f_{\r\s}D_\b\psi_r,\\
\label{ss}s^r_{10}&=\th^{\a\b}\bar\psi_r{\ga_{\a}}^{\r\s} P_L  f_{\b\s}D_\r\psi_r, & s^r_{11}&=\th^{\a\b}\bar\psi_r\ga_\a D_\b D^2\psi_r, & s^r_{12}&=\th^{\a\b}\bar\psi_r{\ga_{\a\b}}^\m D_\m D^2\psi_r.
\end{align}
In the previous expressions we omitted the hats in the background fields and explicit indications of the representations $\rho_r$ to ease the notation. $f_{\mu\nu}$ and ${\cal D}_\alpha f_{\mu\nu}$ are shorthands for  $\rho_r(f_{\mu\nu})$ and $\rho_r({\cal D}_\alpha f_{\mu\nu})$, and will continue being so for the rest of the paper unless otherwise specified. ${\cal D}_\alpha f_{\mu\nu}$ is defined as ${\cal D}_\alpha f_{\mu\nu}=\partial_\alpha f_{\mu\nu}-i[a_\alpha,f_{\mu\nu}]$.
We will use the same basis for U(1) or SU(N), by substituting appropriately the expansions of the U(1) and SU(N) gauge fields given in eq.~\eqref{genexp}  into the definitions for $f_{\m\n}$ and the covariant derivatives of eqs.~\eqref{Sncbos} and \eqref{Sncfer}.
In the non-abelian case, there are other admissible gauge invariant terms, involving symmetric invariant tensors $t^{a_1\dots a_k}$ of the gauge group, such as $\th^{\a\b} \bar\psi_r \gamma^\m t^{a_1\dots a_k}\rho_r(T^{a_1}\dots T^{a_{k-1}})(f_{\a\b})^{a_k} D_\m\psi_r$. The only terms of this type that will appear in the effective action in the SU(N) case involve the combination
\begin{align}\label{Delta}
 \Delta_r^a\equiv d^{abc}\rho_r(T^bT^c),
\end{align}
which, for generic representations, does not belong to the Lie algebra, so that terms constructed with it will be independent of the $s_i$ in eq.~\eqref{ss}. This is not true, for example, for the fundamental and antifundamental representations $F$ and $\bar F$. We will denote by ${\cal F}$ the set of representations $r'$ for which $\Delta^a$ is Lie algebra valued; for these one has
\begin{align}
\label{deltaF}
\Delta_{r'}^a=\frac{N A(r')}{4c_2(r')}\rho_{r'}(T^a),
\end{align}
$c_2(r)$ being the Dynkin index, such that $\Tr_r T^aT^b=c_2(r)\delta^{ab}$. Clearly, for the representations in $\cal F$ no terms other than those in eq.~\eqref{ss} have to be considered. However, for representations that do not satisfy analogous properties, we have to consider the additional terms that follow:
\begin{align}
\nonumber s^r_{13}&=\th^{\a\b}\bar\psi_r\ga^\m P_L \Delta_r^a(f_{\m\b})^aD_\a\psi_r,& s^r_{14}&=\th^{\a\b}\bar\psi_r\ga^\m P_L \Delta_r^a(f_{\a\b})^aD_\m\psi_r,\\
\nonumber s^r_{15}&=\th^{\a\b}\bar\psi_r\ga^\m P_L \Delta_r^a({\cal D}_\m f_{\a\b})^a\psi_r, & s^r_{16}&=\th^{\a\b}\bar\psi_r\ga_\a P_L \Delta_r^a(f_{\b\m})^aD^\m\psi_r,\\
\nonumber
 s^r_{17}&=\th^{\a\b}\bar\psi_r\ga_\a P_L \Delta_r^a({\cal D}^\m f_{\b\m})^a\psi_r, &
s^r_{18}&=\th^{\a\b}\bar\psi_r{\ga_{\a\b}}^\m P_L \Delta_r^a({\cal D}^\n f_{\m\n})^a\psi_r,\\
\nonumber s^r_{19}&=\th^{\a\b}\bar\psi_r{\ga_{\a\b}}^\m P_L  \Delta_r^a(f_{\m\n})^aD^\n\psi_r, & s^r_{20}&=\th^{\a\b}\bar\psi_r{\ga_{\a}}^{\r\s} P_L \Delta_r^a({\cal D}_\b f_{\r\s})^a\psi_r, \\
\label{ss2} s^r_{21}&=\th^{\a\b}\bar\psi_r{\ga_{\a}}^{\r\s} P_L \Delta_r^a( f_{\r\s})^aD_\b\psi_r,& s^r_{22}&=\th^{\a\b}\bar\psi_r{\ga_{\a}}^{\r\s} P_L \Delta_r^a( f_{\b\s})^aD_\r\psi_r.
\end{align}
Finally, we need a basis of terms of order $O(\th)$ with four fermion fields, which will appear when studying the counterterms. It will suffice to consider the following ones
\begin{align}
 \nonumber t^{rs}_1=&\th^{\a\b}(\bar\psi_r\ga_\a P_L \rho_r(T^a)\psi_r)(\bar\psi_s\ga_\b P_L \rho_s(T^a)\psi_s),\,r< s,\\
 \label{ts}t^{rs}_2=&\tilde\th^{\a\b}(\bar\psi_r\ga_\a P_L \rho_r(T^a)\psi_r)(\bar\psi_s\ga_\b P_L \rho_s(T^a)\psi_s),\,r< s,\tilde\theta^{\a\b}=\frac{1}{2}\epsilon^{\a\b\m\n}\th_{\m\n}.
\end{align}
$T^a$ denote the Lie algebra generators ($Y$ in the U(1) case).


\newpage
	\section{Computation of the UV divergences in the effective action involving two fermion fields}
	
	In this section we will proceed to compute the divergent contributions to the effective action involving two fermion fields and no evanescent operators, at one-loop and first order in $\theta$, by calculating the background field 1PI diagrams $\tilde\Gamma^{(n,k)}$ with no external quantum field legs of eq.~\eqref{Gammaexp}, using the Feynman rules in appendix A. Due to the results of ref.~\cite{Martin:2009sg}, the divergent contribution with two fermion fields represents the whole divergent part of the matter sector of the effective action at order $\theta$.

To identify the diagrams that must be computed, we consider the terms with two fermion fields in eqs.~\eqref{ss} and \eqref{ss2}. Doing a similar analysis as that of ref.~\cite{Martin:2009vg}, we have that the terms $s_i$  that may appear in the divergent, non evanescent part of the effective action are of the form $\theta\bar\psi D^3\psi$ --spanned by $s_{11},s_{12}$, which involve at least two fermion fields and have independent two-field contributions-- and $\theta\bar\psi (Df)\psi, \theta\bar\psi f D\psi$, spanned by $s_1$-$s_{10},s_{13}$-$s_{22}$, which involve at least two fermion fields and a gauge field, in such a way that these contributions are independent of each other (with the exception that for representations in which $\Delta_r^a$ of eq.~\eqref{Delta} is Lie algebra valued, $s_{13}$-$s_{22}$ can be expressed in terms of $ s_1$-$s_{10}$). It follows from the previous discussion that the coefficients of the expansion of the divergent contributions to the  effective action involving two fermion fields can be obtained by computing only  1PI diagrams with two fermion fields, $\tilde \Gamma^{(0,1)}$, and with one gauge field and two fermion fields, $\tilde\Gamma^{(1,1)}$.  There is, however, a subtlety: the diagrams with two fermionic legs yield the contribution $-i\tilde \Gamma^{(0,1)}_{ij}\hat{\bar\psi}_i\hat\psi_j$  to the effective action (see eq.~\eqref{Gammaexp}), which fixes the coefficients of $s_{11}$ and $s_{12}$ in the expansion of the effective action in terms of the basis of gauge invariant terms. Then, the diagrams with an external gauge field and two external fermion fields  contribute as  $-i\tilde \Gamma^{(1,1)}_{ij(\m,a)}\hat{\bar\psi}_i\hat\psi_j\hat b_\m^a$ to the effective action, 
which will be a sum of three-field terms coming from both the $s_{11},s_{12}$ combination fixed beforehand and from the $s_1$-$s_{22}$ terms. The three-field contributions  of  $s_{11}$ and $s_{12}$ have to be subtracted in order to get the coefficients of the $s_1$-$s_{22}$ terms.
\subsection{ U(1) case}
\begin{figure}[h]\centering
\includegraphics[scale=1.2]{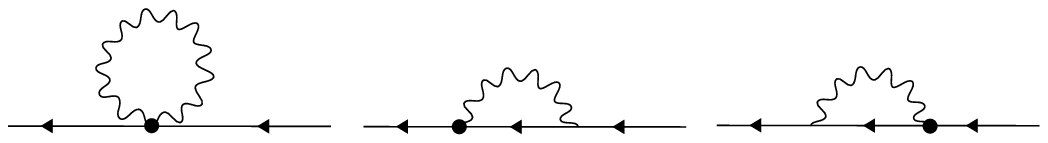}
\caption{Diagrams contributing to  $\tilde\Gamma^{(0,1)}$ at order $\th$ in the U(1) case.}
\label{f:1}
\end{figure}

\begin{figure}[h]\centering
\includegraphics[scale=1.2]{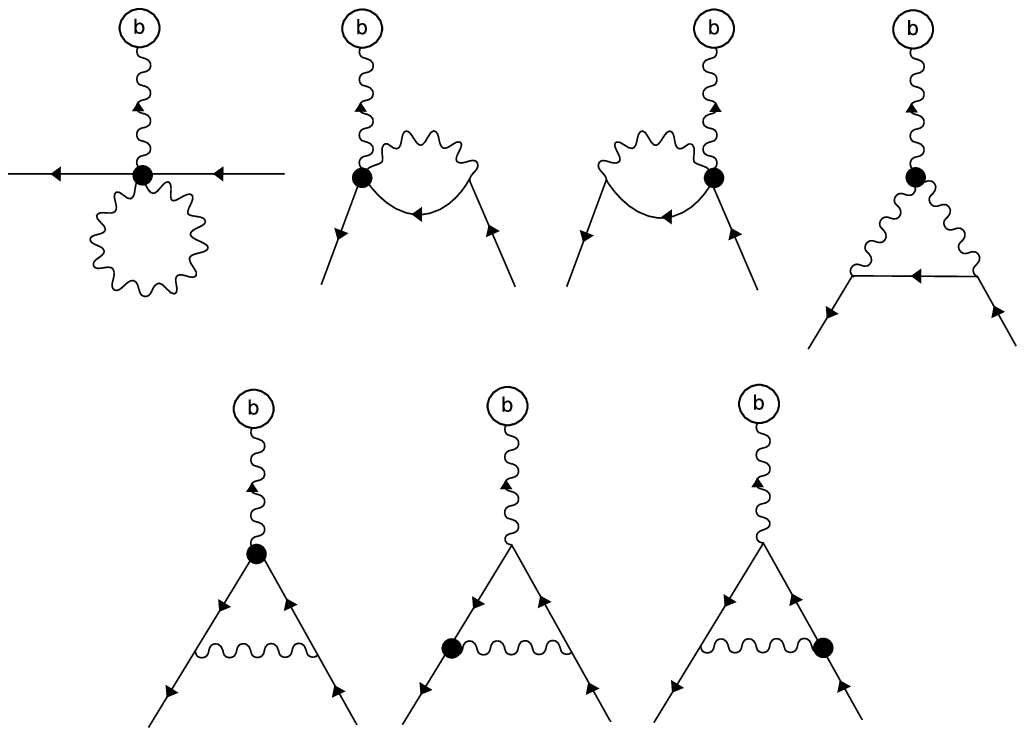}
\caption{Diagrams contributing to  $\tilde\Gamma^{(1,1)}$ at order $\th$ in the U(1) case.}
\label{f:2}
\end{figure}

The diagrams from which one computes the gauge invariant, non-evanescent poles in the effective action at one-loop and order $\th$ are shown in Figs.~\ref{f:1} and \ref{f:2}.

The result of the computations is

\begin{align}\nonumber
&\Gamma^{\rm NC,U(1)}_{\rm div,matter}[b_\m,\psi]=\iDx\Big\{\frac{1}{192\pi^2\epsilon}\th^{\a\b}\sum_r  (e Y_r)^2\,\bar\psi_r\ga_{\a\b\r}P_L D^\r D^2\psi_r\\
\nonumber&+\frac{i}{16\pi^2\epsilon}\sum_r (e Y_r)^3\Big[\Big(\frac{1}{2}-\frac{4x_r}{3}\Big)\th^{\a\b}\bar\psi_r\ga^\m P_L(f_{\m\b})^0D_\a\psi_r+\frac{2x_r}{3}\th^{\a\b}\bar\psi_r\ga^\m P_L(f_{\a\b})^0D_\m\psi_r\\
\nonumber&+\frac{1}{8}\th^{\a\b}\bar\psi_r\ga^\m P_L ({\cal D}_\m f_{\a\b})^0 \psi_r+\Big(\frac{4x_r}{3}-\frac{3}{2}\Big)\th^{\a\b}\bar\psi_r\ga_\a P_L  (f_{\b\m})^0 D^\m\psi_r\\
\nonumber &+\Big(\frac{2x_r}{3}-\frac{3}{4}\Big)\th^{\a\b}\bar\psi_r\ga_\a P_L({\cal D}^\m f_{\b\m})^0  \psi_r+\Big(\frac{1}{24}+\frac{3x_r}{4}\Big)\th^{\a\b}\bar\psi_r{\ga_\a}^{\r\s} P_L ({\cal D}_\b f_{\r\s})^0 \psi_r\\
\label{GammaexpU1}&+\Big(\frac{1}{6}+\frac{3x_r}{4}\Big)\th^{\a\b}\bar\psi_r{\ga_{\a\b}}^{\m} P_L ({\cal D}^\n f_{\m\n})^0 \psi_r+\frac{1}{12}\th^{\a\b}\bar\psi_r{\ga_{\a\b}}^{\m} P_L  (f_{\m\n})^0 D^\n \psi_r\Big]\Big\},
\end{align}
where 
\begin{align*}
x_r\equiv\frac{\tilde e}{e Y_r}
\end{align*}
and the field strength and covariant derivatives are evaluated on the background field, the field strength $(f_{\m\n})^0$  being defined in eq.~\eqref{f0}. We suppressed the hats over the classical fields to ease the notation. 

In terms of the basis of $s_i$ terms in eq.~\eqref{ss}, the divergent contribution to the effective action in eq.~\eqref{GammaexpU1} can be expressed as
\begin{align}\nonumber
\Gamma^{\rm NC,U(1)}_{\rm div,matter}[b_\m,\psi]=&\iDx\sum_r\Big(\frac{(e Y_r )^2i}{16\pi^2\epsilon}\Big[\Big(\frac{1}{2}-\frac{4x_r}{3}\Big)s^r_1+\frac{2x_r}{3}s^r_2+\frac{1}{8}s^r_3+\Big(\frac{4x_r}{3}-\frac{3}{2}\Big)s^r_4\\
&+\Big(\frac{2x_r}{3}-\frac{3}{4}\Big)s^r_5+\Big(\frac{1}{6}+\frac{3x_r}{4}\Big)s^r_6+\frac{1}{12}s^r_7+\Big(\frac{1}{24}+\frac{3x_r}{4}\Big)s^r_8-\frac{i}{12}s_{12}\Big]\Big).
\label{GammadivsU1}
\end{align}


\subsection{SU(N), $\rm N>2$  case}
\begin{figure}[h]\centering
\includegraphics[scale=1.2]{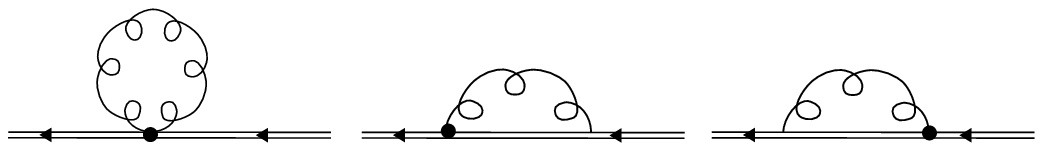}
\caption{Diagrams contributing to  $\tilde\Gamma^{(0,1)}$ at order $\th$ in the SU(N),  $\rm N>2$  case.}
\label{f:3}
\end{figure}

\begin{figure}[h]\centering
\includegraphics[scale=1.2]{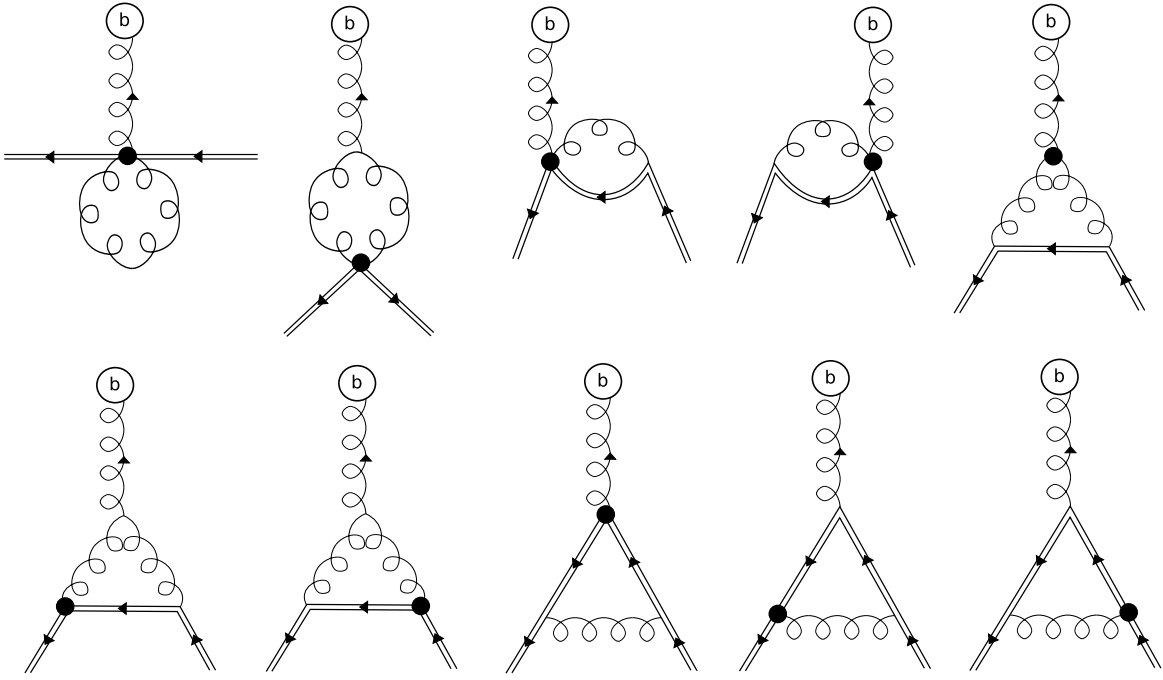}
\caption{Diagrams contributing to  $\tilde\Gamma^{(1,1)}$ at order $\th$ in the SU(N),  $\rm N>2$  case.}
\label{f:4}
\end{figure}

The diagrams from which one computes the gauge invariant, non-evanescent poles in the effective action at one-loop and order $\th$ are shown in Figs.~\ref{f:3} and \ref{f:4}. 

The reconstruction of the divergent contributions to the effective action in terms of a basis of gauge invariant terms is somewhat different in the cases where the fermion multiplet includes representations such as the fundamental or antifundamental in which the $\Delta_r^a$ of eq.~\eqref{Delta} are Lie algebra valued; this is so since, as we already argued, for these representations the terms in eq.~\eqref{ss2} are not independent of those in eq.~\eqref{ss}. Nevertheless, we can write an expression covering all cases as follows,

\begin{align*}\nonumber
&\Gamma^{\rm NC,SU(N)}_{\rm div,matter}[b_\m,\psi]=\iDx\Big\{\frac{g^2 C_2(r)}{192\pi^2\epsilon}\th^{\a\b}\sum_r \bar\psi_r\ga_{\a\b\r}P_L D^\r D^2\psi_r
+\frac{ig^2 N}{16\pi^2\epsilon}\sum_r \Big[\frac{1}{6}\th^{\a\b}\bar\psi_r\ga^\m P_Lf_{\m\b}D_\a\psi_r\\
\nonumber &-\frac{1}{3}\th^{\a\b}\bar\psi_r\ga^\m P_L f_{\a\b}D_\m \psi_r-\frac{1}{8}\th^{\a\b}\bar\psi_r\ga^\m P_L {\cal D}_\m f_{\a\b} \psi_r+\frac{5}{6}\th^{\a\b}\bar\psi_r\ga_\a P_L  f_{\b\m} D^\m\\
\nonumber &+\frac{5}{12}\th^{\a\b}\bar\psi_r\ga_\a P_L{\cal D}^\m f_{\b\m}  \psi_r-\frac{1}{8}\th^{\a\b}\bar\psi_r{\ga_\a}^{\r\s} P_L {\cal D}_\b f_{\r\s} \psi_r-\frac{1}{16}\th^{\a\b}\bar\psi_r{\ga_{\a\b}}^{\m} P_L {\cal D}^\n f_{\m\n} \psi_r\Big]\\
\nonumber &+\frac{ig^2}{16\pi^2\epsilon}\sum_r C_2(r)\Big[\frac{1}{2}\th^{\a\b}\bar\psi_r\ga^\m P_Lf_{\m\b}D_\a\psi_r+\frac{1}{8}\th^{\a\b}\bar\psi_r\ga^\m P_L {\cal D}_\m f_{\a\b} \psi_r-\frac{3}{2}\th^{\a\b}\bar\psi_r\ga_\a P_L  f_{\b\m} D^\m\\
\nonumber &-\frac{3}{4}\th^{\a\b}\bar\psi_r\ga_\a P_L{\cal D}^\m f_{\b\m}  \psi_r+\frac{1}{24}\th^{\a\b}\bar\psi_r{\ga_\a}^{\r\s} P_L {\cal D}_\b f_{\r\s} \psi_r+\frac{1}{6}\th^{\a\b}\bar\psi_r{\ga_{\a\b}}^{\m} P_L {\cal D}^\n f_{\m\n} \psi_r\\
&+\frac{1}{12}\th^{\a\b}\bar\psi_r{\ga_{\a\b}}^{\m} P_L  f_{\m\n} D^\n \psi_r\Big]+\frac{ig^4}{16\pi^2\tilde g^2\epsilon}\sum_{r} \Big[-\frac{2}{3}\th^{\a\b}\bar\psi_r\ga^\m P_L \Delta_r^a(f_{\m\b})^aD_\a\psi_r\\
\nonumber&+\frac{1}{3}\th^{\a\b}\bar\psi_r\ga_\a P_L \Delta_r^a({\cal D}^\m f_{\b\m})^a\psi_r+\frac{2}{3}\th^{\a\b}\bar\psi_r\ga_\a P_L \Delta_r^a(f_{\b\m})^aD^\m\psi_r+\frac{1}{3}\th^{\a\b}\bar\psi_r\ga^\m P_L \Delta_r^a(f_{\a\b})^aD_\m\psi_r\\
\nonumber&+\frac{3}{8}\th^{\a\b}\bar\psi_r{\ga_{\a}}^{\r\s} P_L \Delta_r^a({\cal D}_\b f_{\r\s})^a\psi_r+\frac{3}{8}\th^{\a\b}\bar\psi_r{\ga_{\a\b}}^\m P_L \Delta_r^a({\cal D}^\n f_{\m\n})^a\psi_r\Big]\Big\},
\end{align*}
where all covariant derivatives and field strengths are evaluated on the background field $b_\m$, and we have again suppressed the hats on the classical fields.
In the formulae above, $C_2(r)$ represents the second Casimir of the representation $r$, $C_2(G)$ corresponding to the adjoint representation; it is defined as $\rho_r(T^a T^a)=C_2(r)\mathbb{I}_r$. In terms of the basis $s_i$ of independent, gauge invariant terms of eqs.~\eqref{ss} and \eqref{ss2}, we have, taking into account that $s_{13}$-$s_{20}$ are only independent of the $s_{1}$-$s_{12}$ for representations other than those in the set $\cal F$ for which $\Delta^a_r$ is Lie algebra valued,
\begin{align}
\nonumber&\Gamma^{\rm NC,SU(N)}_{\rm div,matter}[b_\m,\psi]=\iDx\Big(\frac{ig^2N}{16\pi^2\epsilon}\sum_r \Big[\frac{1}{6}s^r_1-\frac{1}{3}s^r_2-\frac{1}{8}s^r_3+\frac{5}{6}s^r_4+\frac{5}{12}s^r_5-\frac{1}{16}s^r_6-\frac{1}{8}s^r_8\Big]\\
\label{GammadivsSUN}
&+\frac{g^2i}{16\pi^2\epsilon}\sum_r C_2(r)\Big[\frac{1}{2}s^r_1+\frac{1}{8}s^r_3-\frac{3}{2}s^r_4-\frac{3}{4}s^r_5+\frac{1}{6}s^r_6+\frac{1}{12}s^r_7+\frac{1}{24}s^r_8-\frac{i}{12}s^r_{12}\Big]\\
\nonumber&+\frac{ig^4}{16\pi^2\tilde g^2\epsilon}\sum_{r\notin \cal F} \Big[-\frac{2}{3}s^r_{13}+\frac{1}{3}s^r_{14}+\frac{2}{3}s^r_{16}+\frac{1}{3}s^r_{17}+\frac{3}{8}s^r_{18}+\frac{3}{8}s^r_{20}\Big]\\
\nonumber&+\frac{ig^4 N}{32\pi^2\tilde g^2\epsilon}\sum_{r\in \cal F}\frac{A(r)}{2c_2(r)}\Big[-\frac{2}{3}s^r_{1}+\frac{1}{3}s^r_{2}+\frac{2}{3}s^r_{4}+\frac{1}{3}s^r_{5}+\frac{3}{8}s^r_{6}+\frac{3}{8}s^r_{8}\Big]\Big).
\end{align}


\section{Computation of the UV divergent contributions in the gauge sector}

In this section we will compute the UV divergences in the bosonic sector that do not involve any evanescent operators and are therefore gauge invariant, using the same strategy employed for the matter sector: choice of gauge $\alpha=1$ and computation of a minimum number of diagrams. To identify the diagrams that need to be computed, we refer to the basis of bosonic gauge invariant terms in eq.~\eqref{rs}. The contributions of the terms $r_1,r_2$ with three gauge fields are independent of each other, and thus, in order to fix their coefficients in the divergent part of the effective action, it suffices to compute the pole of the Green functions involving three background field legs, $\tilde\Gamma^{(3,0)}$ in the notation of eq.~\eqref{Gammaexp}.

These diagrams can have internal gauge or fermionic propagators. We will analyse separately the diagrams with an internal fermionic loop --whose relevant divergences will be shown to vanish-- and with a gauge-field loop.

\subsection{Cancellation of the non-evanescent UV divergences of the fermionic loops}

  The diagrams with internal fermion loops can be analysed with the techniques employed in refs.~\cite{Brandt:2003fx} and \cite{Martin:2007wv}. In these papers it is argued that by appropriately defining the dimensionally regularised interactions, a change of variables can be done in the path integral which is equivalent to inverting the Seiberg-Witten map. This change of variables has unit Jacobian in dimensional regularisation, and allows to compute Green functions with external gauge fields by using interaction vertices in terms of the noncommutative fermions. These diagrams can be computed to all orders in $\theta$ due to the fact that the phase factors generated by the noncommutativity are independent of the loop momenta. As noted in ref.~\cite{Brandt:2003fx}, their UV divergences have zero contributions involving $\epsilon$ tensors, while the vector contributions, computed in ref.~\cite{Martin:2007wv},  are proportional to $\sum_r \Tr_r F_{\m\n}\star F_{\m\n}$, where $F_{\m\n}$ is the noncommutative field strength and $r$ labels the irreducible representations of the fermion fields. Now, expanding $\iDx\sum_r \Tr_r F_{\m\n}\star F_{\m\n}$ with the
Seiberg-Witten map yields an $O(\th)$ contribution to the effective action of the form
\begin{align*}
&\iDx \sum_r\Tr_r F_{\m\n}\star F^{\m\n}=\\
&\big(\sum_r A(r)\big)\iDx \Big(-\frac{1}{4}r_1+r_2\Big)+O(h^2),
\end{align*}
where $r_1$ and $r_2$ are given in eq.~\eqref{rs}, while $A(r)$ are the anomaly coefficients, whose sum $\sum_r A(r)$ must vanish for the theory to be anomaly free. These results apply for both the U(1) and SU(N),  $\rm N>2$  cases. 

The choices of dimensionally regularised interactions that allow for the inversion of the Seiberg-Witten map for fermions differ  by evanescent operators from the choice that was used in the computations of the previous section. Recalling that divergences involving evanescent operators are not physical at one-loop and hence can be ignored, we conclude that the relevant contributions of the fermion loops to the divergences of the diagrams with external gauge field legs are zero. Thus, it only remains to study the divergences involving gauge field loops; we will treat the U(1) and SU(N) cases separately.
\subsection{Gauge-field loop contributions in the U(1) case}
At order $\theta$, there are no one-loop diagrams involving only gauge fields, due to the fact that the commutative action does not have any interaction with three or more gauge fields. Together with the vanishing of the divergences of the fermionic loops, we conclude
\begin{align}\label{GammadivbosU1}
\Gamma^{\rm NC, U(1)}_{\rm div,bos}[b_\m,\psi]=0.
\end{align}
\subsection{Gauge-field loop contributions in the SU(N),  $\rm N>2$  case}
\begin{figure}[h]\centering
\includegraphics[scale=1.2]{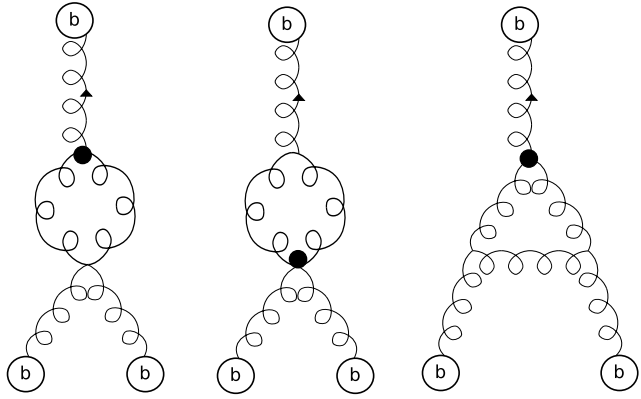}
\caption{Diagrams contributing to  $\tilde\Gamma^{(3,0)}$ at order $\th$ in the SU(N), $\rm N>2$ case.}
\label{f:5}
\end{figure}

The diagrams with three external background field legs involving only internal gauge field propagators are shown in Fig.~\ref{f:5}.
Their associated contribution to the UV divergent part of the effective action is
\begin{align}\label{GammadivbosSUN}
\Gamma^{\rm NC,SU(N)}_{\rm div,bos}[b_\m,\psi]=\iDx\frac{11Ng^2}{48\pi^2\tilde g^2\epsilon}\Big(\frac{1}{8}
r_1-\frac{1}{2}r_2\Big)+O(\th^2)=\frac{11Ng^2}{48\pi^2\epsilon}S^{\rm NC,SU(N)}_{\rm tree,bos}+O(\th^2),
\end{align}
where $S^{\rm NC,SU(N)}_{\rm tree,bos}$ denotes the noncommutative contribution to the bosonic action of eq.~\eqref{SncbosSUN}.

	\section{Studying renormalisability}
	
In the previous sections we calculated the noncommutative UV divergences in both the matter sector  and the bosonic sector. The results are given, expanded in a basis of gauge invariant terms whose elements are displayed in eqs.~\eqref{rs}, \eqref{ss} and \eqref{ss2}, by eqs.~\eqref{GammadivsU1}, \eqref{GammadivsSUN} for the matter sector, and eqs.~\eqref{GammadivbosU1} and \eqref{GammadivbosSUN} for the bosonic sector.

In this section we will check whether these noncommutative divergences  can be subtracted by means of multiplicative renormalisations and the addition of $\theta$-dependent counterterms that vanish on-shell. We will consider separately the U(1) and SU(N) cases.

\subsection{U(1) case}
	
 To start, it is known that the one-loop divergences of QED at order zero in $\theta$ are renormalisable by means of multiplicative renormalisations of fields and parameters. We consider the following form for these multiplicative renormalisations
\begin{align}\label{multiprenU1}
 b_\m&=  Z_b^{1/2}b_\m^{R}, & \psi_r&= (Z_\psi^r)^{1/2}\psi_r^{R},& e&=\mu^{2\epsilon}Z_e e^R, &\tilde e&=\mu^{2\epsilon}dZ_{\tilde e}\tilde e^R, & \theta_{\m\n}&=Z_\th \theta^R_{\m\n},
\end{align}
with $Z_i=1+\delta Z_i$. The gauge invariance present in the calculated divergences forces $\delta Z_e=-\frac{1}{2}\delta Z_b$; the ordinary commutative calculation yields 
\begin{align}
\nonumber\delta Z_b&=-2\delta Z_e=\frac{e^2}{12\pi^2\epsilon},\\
\label{ZsU1}\delta Z^r_\psi&=\frac{e^2 }{16\pi^2\epsilon}.
\end{align}
For the renormalisation of the noncommutative divergences we will also consider $\theta$-dependent counterterms that vanish on-shell. These redundant interactions are of the form
\begin{equation}\label{redundantU1}
\delta S[b,\psi]=\idx \frac{\delta S}{\delta b_\m^0(x)} F_\m^0[b,\psi]+\Big(\sum_r\frac{\delta S}{\delta\psi_{r}(x)}G_r[b,\psi]+{\rm c.c.}\Big),
\end{equation}
which vanish on-shell due to the equations of motion
\begin{align*}
\frac{\delta S[b,\psi]}{\delta b_\m^0(x)}=\frac{\delta S[b,\psi]}{\delta\psi_r(x)}=0.
\end{align*}
In order to preserve gauge symmetry,  $F_\m^0[a,\psi]$ and $G_r[a,\psi]$ have to transform in four dimensions, under  a U(1)  gauge transformation with gauge parameter $\lambda=e\lambda^0 Y$, as
 \begin{align*}
s F^0_\m=0,\quad sG_r=ie\lambda^0 \rho_r(Y) G_r.
\end{align*}
We consider the following solutions
\begin{align}
\nonumber\delta F_\m^0[b,\psi]=&y_1\th^{\a\b}({\cal D}_\m f_{\a\b})^0+y_2{\th_\m}^\a({\cal D}^\n f_{\n\a})^0+\sum_r y^r_3{\th_\m}^\a\bar\psi_r\ga_\a P_L\psi_r+i\sum_r y^r_4{\th}^{\a\b}\bar\psi_r\ga_{\m\a\b} P_L\psi_r\\
\label{FGU1}&+y_5{\tilde{\th}_\m}^{\,\,\,\,\b} (D^\n f_{\n\b})^0,\,\,y_i\in\mathbb{R},\\
\nonumber G_r[b,\psi]=&k^r_1\th^{\a\b}f_{\a\b}P_L\psi_r+k^2_r\th^{\a\b}{\ga_{\a\m}}P_L{f_\b}^\m \psi_r+k^r_3\th^{\a\b}\ga_{\a\m}P_L D_\b D^\m\psi_r+k^r_4\th^{\a\b}\ga_{\a\b}P_L D^2\psi_r\\
\nonumber&+k^r_5\tilde{\th}^{\a\b}\ga_5 P_L f_{\a\b}\psi_r,\,\,k_i\in\mathbb{C}.
\end{align}
Note that we used $f_{\m\n}=e Y (f_{\m\n})^0, D_\r f_{\m\n}=eY (D_\r f_{\m\n})^0$ in $G_r$. The counterterm action that results at $O(h)$ from the multiplicative renormalisations of eq.~\eqref{multiprenU1} and the unphysical $\theta$-dependent counterterms of eqs.~\eqref{redundantU1} and \eqref{FGU1} has the following expansion in the basis of eqs.~\eqref{rs}, \eqref{ss} and \eqref{ts} particularised for the U(1) theory
\begin{align}
\nonumber S^{\rm NC,U(1)}_{\rm ct}&=S^{\rm NC,U(1)}_{\rm ct,bos}+S^{\rm NC,U(1)}_{\rm nc,matter},\\
\label{SctU1}
 S^{\rm NC,U(1)}_{\rm ct,bos}&=\idx\sum_{i}B_ir_i,\quad S^{\rm NC,U(1)}_{\rm ct,matter}
=i\idx\sum_{r,i} C_i^r s^r_i+\idx\sum_{i,r< s}D_i^{rs} t^{rs}_i,
\end{align}
\begin{align*}
B_1&=\frac{\tilde e}{16}\Big(\delta Z_{\tilde e}+\delta Z_\th+\frac{3}{2}\delta Z_b\Big), & B_2&=-\frac{\tilde e}{4}\Big(\delta Z_{\tilde e}+\delta Z_\th+\frac{3}{2}\delta Z_b\Big),\\
C^r_1&=\frac{1}{2}(\delta Z_\th\!+\!\delta Z_\psi)\!-\!(k^r_2)^*\!-\!k^r_2, & 
 C^r_2&=-\frac{1}{4}(\delta Z_\th\!+\!\delta Z_\psi)+(k^r_1)^*+k^r_1+\!\frac{i}{2}(k^r_3)^*\!-\!\frac{i}{2}k^r_3,\\
C^r_3&=-i y_1+k^r_1-\frac{1}{2}k^r_2+\frac{i}{2}(k^r_3)^*,&
 C^r_4&=-(k^r_2)^*-k^r_2-i(k^r_3)^*-ik^r_3-4i(k^r_4)^*,\\
C^r_5&=-2i(k^r_4)^*-k^r_2+iy_2-\frac{i}{e Y_r} y^r_3, &
C^r_6&=-\frac{1}{eY_r}y^r_4-\frac{1}{2}y_7-i(k^r_4)^*-ik^r_5,\\
C^r_7&=-2i(k^r_4)^*-i(k^r_5)^*-ik^r_5, &
 C^r_8&=-\frac{1}{2}k^r_2+\frac{i}{2}(k^r_3)^*+\frac{i}{2}k^r_3-ik^r_5,\\
C^r_9&=\frac{i}{2}(k^r_3)^*+\frac{i}{2}k^r_3-ik^r_5-i(k^r_5)^*, &
 C^r_{10}&=(k^r_2)^*-k^r_2+i(k^r_3)^*+ik^r_3,\\
C^r_{11}&=-(k^r_3)^*-k^r_3-2(k^r_4)^*-2k^r_4,&
 C^r_{12}&=-(k^r_4)^*+k^r_4,\\
D^{rs}_1&=e\Big(\frac{y^s_3}{Y_s}-\frac{y^r_3}{Y_r}\Big), &
D^{rs}_2&=2e\Big(\frac{y^s_4}{Y_s}-\frac{y^r_4}{Y_r}\Big).
\end{align*}
The counterterm action above must cancel the divergences in the bosonic and matter sectors. This means
\begin{align}
 S^{\rm NC,U(1)}_{\rm ct,bos}=-\Gamma^{\rm NC, U(1)}_{\rm div,bos},\quad S^{\rm NC,U(1)}_{\rm ct,matter}=-\Gamma^{\rm NC, U(1)}_{\rm div,matter}\label{eqcount}
\end{align}
Since we expanded both the divergences and counterterms in a basis of independent terms $r_i,s_i,t_i$, in order to solve eq.~\eqref{eqcount} it suffices to match the real and imaginary parts of the coefficients of these expansions, which are given in eqs.~\eqref{GammadivsU1}, \eqref{GammadivbosU1} and \eqref{SctU1}. As a result, we get the following system of equations:

\begin{align}\nonumber
r_1,r_2&: & &\delta Z_{\tilde e}+\delta Z_\th+\frac{3}{2}\delta Z_b=0,\\
\nonumber s^r_1&:  & &\frac{1}{2}(\delta Z\th+\delta Z_\psi)-2{\rm Re}k^r_2=-\frac{(e Y_r)^2}{16\pi^2\epsilon}\Big(\frac{1}{2}-\frac{4 x_r}{3}\Big),\\
\nonumber s^r_2&:  &&-\frac{1}{4}(\delta Z_\th+\delta Z_\psi)+2{\rm Re}k^r_1+{\rm Im}k^r_3=-\frac{(e Y_r)^2 x_r}{24\pi^2\epsilon},\\
\nonumber s^r_3&: &&y_1-\frac{1}{2}{\rm Re}k^r_3-{\rm Im}k^r_1+\frac{1}{2}{\rm Im}k^r_2=0, \quad \frac{1}{2}{\rm Im}k^r_3+{\rm Re}k^r_1-\frac{1}{2}{\rm Re}k^r_2=-\frac{(e Y_r)^2}{128\pi^2\epsilon},\\
\nonumber s^r_4&: && {\rm Re}k^r_3+2{\rm Re}k^r_4=0,\quad -2{\rm Re}k^r_2-4{\rm Im}k^r_4=-\frac{(eY_r)^2}{16\pi^2\epsilon}\Big(-\frac{3}{2}+\frac{4 x_r}{3}\Big), \\
\nonumber s^r_5&: && eY_r(-y_2+2{\rm Re}k^r_4+{\rm Im}k^r_2)+y^r_3=0,\quad
-{\rm Re}k^r_2-2{\rm Im}k^r_4=-\frac{(eY_r)^2}{16\pi^2\epsilon}\Big(-\frac{3}{4}+\frac{2 x_r}{3}\Big),  \\
\nonumber s^r_6&: && {\rm Re}k^r_4+{\rm Re}k^r_5=0,\quad
-\frac{y^r_4}{eY_r}-\frac{1}{2}y_7-{\rm Im}k^r_4+{\rm Im}k^r_5=-\frac{(eY_r)^2}{16\pi^2\epsilon}\Big(\frac{1}{6}+\frac{3x_r}{4}\Big), \\
\nonumber s^r_7&: && {\rm Re}k^r_4+{\rm Re}k^r_5=0,\quad-2{\rm Im}k^r_4=-\frac{(eY_r)^2}{192\pi^2\epsilon}, \\
\nonumber s^r_8&:&& -{\rm Re}k^r_3+{\rm Re}k^r_5+\frac{1}{2}{\rm Im}k^r_2=0,\quad{\rm Im}k^r_5-\frac{1}{2}{\rm Re}k^r_2=-\frac{(eY_r)^2}{16\pi^2\epsilon}\Big(\frac{1}{24}+\frac{3x_r}{4}\Big), \\
\nonumber s_9^r&: && -{\rm Re}k^r_3+2{\rm Re}k^r_5=0,\\
\nonumber s^r_{10}&: && {\rm Im}k^r_2-{\rm Re}k^r_3=0,\\
\nonumber s^r_{11}&: && -{\rm Re}k^r_3-2{\rm Re}k^r_4=0,\\
 \nonumber s^r_{12}&: && -2{\rm Im}k^r_4=-\frac{(eY_r)^2}{192\pi^s\epsilon},\\
t^{rs}_1,t^{rs}_2&: && \frac{y^s_3}{Y_s}-\frac{y^r_3}{Y_r}=0=\frac{y^s_4}{Y_s}-\frac{y^r_4}{Y_r}.\label{systemU1}
\end{align}
The equations are compatible; we find the following solutions (making use of eq.~\eqref{ZsU1}):
\begin{align}
\nonumber \d Z_\th&=\frac{e^2}{3\pi^2\epsilon}(x_r-1), & \d Z_{\tilde e}&=\frac{e^2}{24\pi^2\epsilon}(5-8x_r),\\
\nonumber y^r_3&= cY^r, c\in \mathbb{R}, & y^r_4&=c'Y^r, c'\in \mathbb{R},\\
\nonumber y_1&={\rm Im}k^r_1,& y_2&=\frac{c}{e},\\
\label{solutionU1}y_7&=-\frac{2}{e}c'+\frac{e^2}{24\pi^2\epsilon}(x_r-1), & {\rm Re}k^r_1&=-\frac{1}{2}{\rm Im}k^r_3+\frac{e^2}{384\pi^2\epsilon}(8x_r-13), \\
 \nonumber{\rm Im}k^r_5&=-\frac{e^2}{384\pi^2\epsilon}(10x_r+11 ),& {\rm Im}k^r_4&=\frac{e ^2}{384\pi^2\epsilon},\\
\nonumber {\rm Re}k^r_2&=\frac{e^2}{96\pi^2\epsilon}(4x_r-5),& {\rm Im}k^r_2&={\rm Re}k^r_3=2{\rm Re}k^r_5=-2{\rm Re}k^r_4.
\end{align}
The couplings $\tilde e$ and $\theta$, as well as $y_1,y_2,y_7$, must be independent of the representations $r$ of the fermions. Imposing this on the solutions of eq.~ \eqref{solutionU1}, we are forced to demand
\begin{align*}
x_r=\frac{\tilde e}{e Y_r}=0\Rightarrow \tilde e=0.
\end{align*}
I.e., renormalisability is attainable by fixing the trace ambiguity in the tree-level bosonic action in such a way that the coupling $\tilde e$ of eq.~\eqref{gtilde} is equal to zero. When $\tilde e=0$, the noncommutative tree-level contributions in the bosonic sector vanish, as well as the bosonic divergences and the counterterms along $r_1,r_2$; then the equations in eq.~\eqref{systemU1} coming from the coefficients along $r_1,r_2$, as well as the solution for $\delta Z_{\tilde e}$ in eq.~\eqref{solutionU1} can be ignored. Thus, the choice $\tilde e=0$ is consistent at the one-loop level and at $O(\th)$. 


\subsection{SU(N),  $\rm N>2$  case}

In the SU(N) case we consider the multiplicative renormalisation of fields and parameters that follow
\begin{align*}
 b_\m&=  Z_b^{1/2}b_\m^{R}, & \psi_r&= (Z^r_\psi)^{1/2}\psi_r^{R},& g&=\mu^{-\epsilon}\,Z_g g^R,& \tilde g&=\mu^{-\epsilon}\, Z_{\tilde g}\tilde g^R, & \theta_{\m\n}&=Z_\th \theta^R_{\m\n},
\end{align*}
Gauge invariance demands now $\delta Z_b=0$, and the $O(\th^0)$ calculation yields the following results
\begin{align}\label{ZsSUN}
\delta Z^r_\psi&=\frac{g^2 C_2(r)}{16\pi^2\epsilon},\\
\nonumber
\delta Z_g&=\frac{g^2}{16\pi^2\epsilon}\Big[\frac{11}{6}C_2(G)-\frac{2}{3}\sum_r c_2(r)\Big].
\end{align}
As before, we consider $\th$-dependent counterterms that vanish on-shell, 
of the form
\begin{equation}\label{redundantSUN}
\delta S[b,\psi]=\idx \frac{\delta S}{\delta b_\m^a(x)} F_\m^a[b,\psi]+\Big(\sum_r\frac{\delta S}{\delta\psi_{r}(x)}G_r[b,\psi]+{\rm c.c.}\Big),
\end{equation}
where now gauge invariance demands --defining $F_\m=F_\m^aT^a$--
 \begin{align*}
s F_\m=-i[F_\m,\la],\quad sG_r=i\lambda G_r,
\end{align*}
and we consider the following solutions
\begin{align}
\nonumber F_\m=&y_1\th^{\a\b}{\cal D}_\m f_{\a\b}+y_2{\th_\m}^\a{\cal D}^\n f_{\n\a}+\sum_r y^r_3{\th_\m}^\a(\bar\psi_r\ga_\a P_L T^a\psi_r)T^a\\
\nonumber &+i\sum_r y^r_4{\th}^{\a\b}(\bar\psi_r\ga_{\m\a\b} P_L T^a\psi_r)T^a+y_5{\tilde{\th}_\m}^{\,\,\,\,\b} D^\n f_{\n\b}, \,\,y_i\in\mathbb{R},\\
\nonumber G_{r,L}=&k^r_1\th^{\a\b}f_{\a\b}P_L\psi_r+k^2_r\th^{\a\b}{\ga_{\a\m}}P_L{f_\b}^\m \psi_r+k^r_3\th^{\a\b}\ga_{\a\m}P_L D_\b D^\m\psi_r+k^r_4\th^{\a\b}\ga_{\a\b}P_L D^2\psi_r\\
&+k^r_5\tilde{\th}^{\a\b}\ga_5 P_L f_{\a\b}\psi_r,\,\,k_i\in\mathbb{C}.
\label{redundant2}
\end{align}
If there are representations $r\notin{\cal F}$ for which the $\Delta_r^a$ of eq.~\eqref{Delta} are not Lie algebra valued, we can consider additional independent contributions to $F_\m,G_{r,L}$, involving $\Delta_r^a$. These are 
\begin{align*}
\nonumber G'_{r,L}=&k^r_6\th^{\a\b}\Delta_r^a(f_{\a\b})^aP_L\psi_r+k^r_7\th^{\a\b}{\ga_{\a\m}}P_L\Delta_r^a({f_\b}^\m)^a \psi_r+k^r_8\tilde{\th}^{\a\b}\ga_5 P_L \Delta_r^a(f_{\a\b})^a\psi_r; \,\,\,k_i\in\mathbb{C},
\end{align*}

In terms of the elements in the basis of eqs.~\eqref{rs}, \eqref{ss}, \eqref{ss2} and \eqref{ts}, the $O(\th)$ counterterm action that results from the multiplicative renormalisations of eq.~\eqref{ZsSUN} and the $\th$-dependent counterterms that result from eqs.~\eqref{redundantSUN} and \eqref{redundant2} is, using the same notation as in the U(1) case, the following
\begin{align}
\nonumber S^{\rm NC,SU(N)}_{\rm ct}&=S^{\rm NC,SU(N)}_{\rm ct,bos}+S^{\rm NC,SU(N)}_{\rm ct,matter},\\
\nonumber
 S^{\rm NC,SU(N)}_{\rm ct,bos}&=\idx\sum_{i}B_ir_i,\\
\label{SctSUN}S^{\rm NC,SU(N)}_{\rm ct,matter}
=&i\idx\sum_{r,i<13} C_i^r s^r_i+i\idx\sum_{r'\notin{\cal F},i>12} C_i^{r'} s^{r'}_i+\idx\sum_{r<s,i}D_i^{rs} t^{rs}_i,
\end{align}
\begin{align*}
B_1&=\frac{1}{8\tilde g^2}\Big(-2\delta Z_{\tilde g}+\delta Z_\th\Big), & B_2&=-\frac{1}{2\tilde g^2}\Big(-2\delta Z_{\tilde g}+\delta Z_\th\Big),\\
C^r_1&=\frac{1}{2}(\delta Z_\th+\delta Z_\psi)-(k^r_2)^*-k^r_2, & C^r_2&=-\frac{1}{4}(\delta Z_\th+\delta Z_\psi)+(k^r_1)^*+k^r_1+\frac{i}{2}(k^r_3)^*-\frac{i}{2}k^r_3,\\
C^r_3&=-i y_1+k^r_1-\frac{1}{2}k^r_2+\frac{i}{2}(k^r_3)^*, & C^r_4&=-(k^r_2)^*-k^r_2-i(k^r_3)^*-ik^r_3-4i(k^r_4)^*,\\
C^r_5&=iy_2-\frac{i}{2g^2} y^r_3-2i(k^r_4)^*-k^r_2, & C^r_6&=-\frac{1}{2g^2}y^r_4-\frac{1}{2}y_5-i(k^r_4)^*-ik^r_5,\\
C^r_7&=-2i(k^r_4)^*-i(k^r_5)^*-ik^r_5, & C^r_8&=-\frac{1}{2}k^r_2+\frac{i}{2}(k^r_3)^*+\frac{i}{2}k^r_3-ik^r_5,\\
C^r_9&=\frac{i}{2}(k^r_3)^*+\frac{i}{2}k^r_3-ik^r_5-i(k^r_5)^*, & C^r_{10}&=(k^r_2)^*-k^r_2+i(k^r_3)^*+ik^r_3,\\
C^r_{11}&=-(k^r_3)^*-k^r_3-2(k^r_4)^*-2k^r_4,& C^r_{12}&=-(k^r_4)^*+k^r_4,\\
C^{r'}_{13}&=-(k^{r'}_7)^*-k^{r'}_7, & C^{r'}_{14}&=(k^{r'}_6)^*+k^{r'}_6,\\
C^{r'}_{15}&=k^{r'}_6-\frac{1}{2}k^{r'}_7, & C^{r'}_{16}&=-(k^{r'}_7)^*-k^{r'}_7,\\
C^{r'}_{17}&=-k^{r'}_7, & C^{r'}_{18}&=-ik^{r'}_8,\\
C^{r'}_{19}&=-i(k^{r'}_8)^*-ik^{r'}_8, & C^{r'}_{20}&=-\frac{1}{2}k^{r'}_7-ik^{r'}_8,\\
C^{r'}_{21}&=-ik^{r'}_8-i(k^{r'}_8)^*, & C^{r'}_{22}&=(k^{r'}_7)^*-k^{r'}_7,\\
D^{rs}_1&=y^s_3-y^r_3, &
D^{rs}_2&=2y^s_4-2y^r_4.
\end{align*}
To demand renormalisability, we impose
\begin{align}\label{counteqSUN}
 S^{\rm NC,SU(N)}_{\rm ct,bos}=-\Gamma^{\rm NC, SU(N)}_{\rm div,bos},\quad S^{\rm NC,SU(N)}_{\rm ct,matter}=-\Gamma^{\rm NC, SU(N)}_{\rm div,matter}.
\end{align}
Matching the coefficients of the expansion of the terms in eq.~\eqref{counteqSUN} in the basis of elements $r_i$, $s^r_i$ and $t_i^{rs}$ --where the expansions are given in eqs.~\eqref{GammadivsSUN}, \eqref{GammadivbosSUN} and \eqref{SctSUN}-- we 
obtain a system of equations.

Let's consider first the case in which there are fermions belonging to representations $r'\notin{\cal F.}$  Matching the coefficients of the $s^{r'}_{13}$ and $s^{r'}_{16}$, we get the following equations
\begin{align*}
& s^{r'}_{13}:\,\,\,{\rm Re}k^{r'}_7=-\frac{g^4N}{96\pi^2\tilde g^2\epsilon}, & s^{r'}_{16}:\,\,\,{\rm Re}k^{r'}_7=\frac{g^4N}{96\pi^2\tilde g^2\epsilon}, 
\end{align*}
which are clearly incompatible unless
\begin{align*}
 \frac{1}{\tilde g^2}=0.
\end{align*}
For $\frac{1}{\tilde g^2}=0$, the divergences along the terms $s^{r'}_{13}$-$s^{r'}_{22}$ in eq.~\eqref{GammadivsSUN} vanish, and the corresponding equations for the counterterms of eq.~\eqref{counteqSUN} along $s^{r'}_{13}$-$s^{r'}_{22}$ can be solved by demanding
\begin{align*}
 k^{r'}_6=k^{r'}_7=k^{r'}_8=0.
\end{align*}
It remains to solve the equations for the coefficients of the terms along $r_i$ and $s^r_{1}$-$s^r_{13}$, for arbitrary representations $r$, belonging or not to $\cal F$. From eq.~\eqref{GammadivsSUN}, it is clear that for $\frac{1}{\tilde g^2}=0$ the matching equations for the counterterms will be the same for all irreducible representations $r$. As  was done in the U(1) case, these equations can be obtained straightforwardly from eqs.~\eqref{GammadivsSUN}, \eqref{GammadivbosSUN}, \eqref{SctSUN} and \eqref{counteqSUN}; for the sake of brevity we will not display them here, but we will note that they are compatible  --but for the identities coming from the matching along $r_1,r_2$, they are the same equations that were found for anomaly safe GUT theories in ref.~\cite{Martin:2009vg}-- and have the following solutions
\begin{align}
\nonumber y_1&={\rm Im}k^r_1,& y^r_3&=c\in\mathbb{R},\\
\nonumber y^r_4&=-y_5 g^2-\frac{g^4}{384\pi^2\epsilon}(16 C_2(r)-13 N),&
 \delta Z_\th&=-\delta Z_\psi-\frac{g^2}{48\pi^2\epsilon}(13C_2(r)-4N), \\
 \nonumber {\rm Re}k^r_1&=-\frac{1}{2}{\rm Im}k^r_3-\frac{g^2}{384\pi^2\epsilon}(13C_2(r)-8N), & {\rm Im}k^r_5&=-\frac{g^2}{384\pi^2\epsilon}(11C_2(r)-8N),\\
 \nonumber{\rm Im}k^r_4&=\frac{g^2 C_2(r)}{384\pi^2\epsilon},& {\rm Re}k^r_2&=-\frac{5g^2}{192\pi^2\epsilon}(2C_2(r)-N),\\
\nonumber {\rm Im}k^r_2&={\rm Re}k^r_3=2{\rm Re}k^r_5=-2{\rm Re}k^r_4, & \delta Z_{\tilde g}&=\frac{1}{2}\delta Z_\th+\frac{11g^2}{96\pi^2\epsilon},\\
y_2&=\frac{1}{2g^2}c, &y_4^r&=c'\in\mathbb{R}.\label{solutionsSUN1}
\end{align}
It should be recalled that  $y_1,y_2,y_5$ and $\delta Z_\th$ must be flavour independent  (see eq.~\eqref{redundant2}).  Imposing this in the previous solutions, one must require that all flavours have identical  $C_2(r)$; this can be achieved without violating the anomaly cancellation conditions by considering all fields in a single representation plus its conjugate. 

Next, we should analyse the case in which there are no fermion fields in representations other than those belonging to ${\cal F}$, as when the anomaly-free matter content is given by a multiplet in the fundamental and another one in the antifundamental. In this case, defining
\begin{align}\label{defz}
z_r\equiv\frac{A(r)g^2}{2\tilde g^2c_2(r)},
\end{align}
the terms $s_j,j>13$ of eq.~\eqref{ss2} are not independent of the terms $s_i,i<13$; it can be readily seen from eqs.~\eqref{deltaF} that one has
\begin{align*}
 s^r_{j}=\frac{\tilde g^2 N z_r}{2g^2}s^r_{j-12},\,j=13,\dots,22,
\end{align*}
with $z_r$ as in eq.~\eqref{defz}. Using the previous identity applied to eqs.~\eqref{GammadivsSUN}, \eqref{GammadivbosSUN}, \eqref{SctSUN} and \eqref{counteqSUN}, one gets a system of equations involving the terms $r_i,s_i,i=1,\cdots 12,t_i$ of eqs.~\eqref{rs}, \eqref{ss} and \eqref{ts}. The system is again compatible and has as solutions
\begin{align}\nonumber
y_1&={\rm Im}k^r_1,& y^r_3&=c\in\mathbb{R},\\
\nonumber y^r_4&=\!-y_5 g^2\!+\!\frac{g^4}{384\pi^2\epsilon}(N(4z_r\!+13)\!-\!16 C_2(r)),&
 \delta Z_\th&=\!-\delta Z_\psi\!+\!\frac{g^2}{48\pi^2\epsilon}(4N(1\!+\!z_r)\!-\!13C_2(r)), \\
\nonumber  {\rm Re}k^r_1&=\!-\frac{1}{2}{\rm Im}k^r_3\!+\!\frac{g^2}{384\pi^2\epsilon}(2N(z_r\!+4)\!-\!13C_2(r)), & {\rm Im}k^r_5&=-\frac{g^2}{768\pi^2\epsilon}(22C_2(r)+N(5z_r-16)),\\
 \nonumber {\rm Im}k^r_4&=\frac{g^2 C_2(r)}{384\pi^2\epsilon},& {\rm Re}k^r_2&=\frac{g^2}{192\pi^2\epsilon}(N(2z_r+5)-10C_2(r)),\\
 \nonumber {\rm Im}k^r_2&={\rm Re}k^r_3=2{\rm Re}k^r_5=-2{\rm Re}k^r_4,& \delta Z_{\tilde g}&=\frac{1}{2}\delta Z_\th+\frac{11g^2}{96\pi^2\epsilon},\\
y_2&=\frac{1}{2g^2}c, &y_4^r&=c'\in\mathbb{R}.\label{solutionsSUN2}
\end{align}
Again, we have to demand the independence of $y_1,y_2,y_5$ and $\delta Z_\th$ of the representation $r$. Since $z(r)$, defined in eq.~\eqref{defz}, involves $A(r)/c_2(r)$, which will differ in the representations considered --$c_2(r)$ is positive, while some $A(r)$ will have necessarily different signs to make the cancellation of the anomaly  possible-- we must have $z(r)=0$ by imposing again
\begin{equation*}
 \frac{1}{\tilde g^2}=0.
\end{equation*}
After doing this, the solutions of eq.~\eqref{solutionsSUN2} are equivalent to those of eq.~\eqref{solutionsSUN1}.

From the previous results, it is clear that the SU(N), $\rm N>2$  theory is one-loop renormalisable only when the trace ambiguity is fixed in such a way that $ (\tilde g_{\rm tree})^{-2}=0$ and when ordinary the matter content is vectorial, consisting of two multiplets in representations that are conjugate of each other; the choice of representation remains arbitrary. Note that when $(\tilde g_{\rm tree})^{-2}=0$, the noncommutative tree-level contributions, the counterterms and the divergences in the bosonic sector vanish, and then the equations for the counterterms along $r_1,r_2$, as well as the solutions for $\delta Z_{\tilde g}$ of eqs.~\eqref{solutionsSUN1} and \eqref{solutionsSUN2}, can be ignored. Hence, the condition $(\tilde g_{\rm tree})^{-2}=0$ survives the renormalisation procedure.


	\section{Discussion}
	
  In this paper we have computed, at one-loop and first order in the noncommutativity parameters $\theta$, the UV divergent contributions involving zero or two fermion fields and an arbitrary number of gauge fields, to the background-field effective action of two types of noncommutative gauge theories: the GUT compatible formulation of QED, and SU(N),  $\rm N>2$  GUTs with an arbitrary, anomaly-free matter content. We used the framework of BMHV dimensional regularisation, neglecting contributions to the divergences involving operators that vanish in the limit $D\rightarrow4$, which are unphysical at one-loop. The results can be summarised as follows: when the trace ambiguities of the models are fixed in such way that the $O(\th)$ triple gauge field couplings vanish at tree level, the off-shell divergences of both the U(1) and SU(N) theories can be renormalised by means of multiplicative renormalisations of fields and couplings --including the noncommutativity parameters $\th$-- and the addition of unphysical $\th$-dependent counterterms that vanish on-shell. In the SU(N) case, this is possible whenever the irreducible representations in the matter multiplet share the second Casimir. When the previous conditions are met, given the results of ref.~\cite{Martin:2009sg} concerning the vanishing of 4 fermion UV divergences in noncommutative GUT compatible theories, the full one-loop, $O(\th)$ effective action is renormalisable.

  The restriction to theories in which the $O(\th)$ triple gauge field couplings vanish  --recall that these interactions are proportional to $\tilde e,\frac{1}{\tilde g^2}$, as shown in eq.~\eqref{Sbos}-- constrains the ambiguity in the trace of the bosonic degrees of freedom. Looking at the expressions for the couplings $\tilde e,\frac{1}{\tilde g^2}$ in eq.~\eqref{gtilde}, their vanishing can be achieved by summing over all the matter representations with a common coefficient, as a consequence of anomaly cancellation conditions. Thus, the choice $\tilde e=\frac{1}{\tilde g^2}=0$ can be regarded as the simplest or minimal one, dictated only by the matter content of the theory, as was already pointed out in ref.~\cite{Aschieri:2002mc}. Our calculations show that, once $\tilde e=\frac{1}{\tilde g^2}=0$ are fixed to zero at tree-level, there is no need to renormalise these parameters due to the vanishing of the bosonic divergences, so that the choice is consistent with one-loop, $O(\th)$ quantum corrections. As was emphasised in the introduction, the only physical parameters of the theory are the coupling constants $e,g$ and the noncommutativity parameters $\theta$. The off-shell renormalisation only required the addition of counterterms arising either from multiplicative renormalisation of the tree-level parameters and fields, or depending on unphysical couplings. Renormalisable S-Matrix elements --which are insensitive to field renormalisations and the on-shell vanishing counterterms-- can be obtained by considering only the multiplicative renormalisation of these physical parameters. Thus, the one-loop, order $\theta$ renormalisation procedure did not require an enlargement of the set of physical couplings; this is essential in raising hopes that the well-behaved theories found in this paper might be indeed renormalisable in the proper sense -meaning that they keep yielding sensible S-Matrix elements that only depend on a finite number of physical couplings- when higher order corrections are taken into account. 

 For $\tilde e=\frac{1}{\tilde g^2}=0$ the theories are similar to the anomaly safe GUTs analysed in ref.~\cite{Martin:2009vg}, and the results too. However, two important distinctions should be made: in the theories analysed in this paper, the anomaly cancellation conditions impose constraints on the allowed matter representations, and, on the other hand, the fact that the choice $\tilde e=\frac{1}{\tilde g^2}=0$ survives the renormalisation procedure is nontrivial, coming from the vanishing of the fermionic contributions to the UV divergences in the bosonic sector: for anomaly safe theories, the contributions to the UV divergences of the form $\Tr\th fff$ vanish automatically, while in the U(1) and SU(N),  $\rm N>2$  case they vanish because their coefficient turns out to be proportional to the total anomaly. In the SU(N) case, the condition that the representations must share the same second Casimir, added to the condition of anomaly freedom, demands that the ordinary matter sector has to be nonchiral, in the sense that it must consist of two multiplets in representations that are conjugate of each other; the noncommutative interactions, since they cannot be written in terms of noncommutative Dirac fermions, remain chiral. Thus, our results overcome the nonrenormalisability observed for non-GUT compatible formulations of noncommutative QED and SU(N) gauge theories with ordinary vector matter, whose matter sectors have unrenormalisable divergences. This increases the number of noncommutative gauge theories in the enveloping algebra approach with fermionic matter which have been shown to be one-loop renormalisable at order $\theta$, and further motivates the study of noncommutative GUT compatible theories. The list of one-loop, $O(\th)$ renormalisable noncommutative theories so far includes noncommutative SU(N) super Yang-Mills, anomaly safe GUTs, and the renormalisable theories found in this paper. It should be pointed out that due to the fact that renormalisable SU(N) GUT theories are nonchiral, noncommutative embeddings of the Standard Model into renormalisable anomaly safe GUT theories should be preferred over embeddings into SU(N) GUTs. Nevertheless, when modelling the strong interactions alone, a GUT compatible SU(N) model stands out as the simplest renormalisable choice at one-loop and order $\theta$. 

  Of course, even though we mentioned Standard Model embeddings, our results about renormalisability apply for theories without Higgs sectors. It is an open and very interesting problem to study the renormalisability of GUT compatible theories with a Higgs sector. On the other hand, our results apply to GUT compatible theories with simple groups; it is still pending to analyse the case of semisimple groups, such as the GUT compatible Standard Model of ref.~\cite{Aschieri:2002mc}. Even though renormalisability could demand similar restrictions for the matter content as those found in this paper, such as common second Casimirs for the different multiplets, this would be satisfied by fields in the (anti)fundamental representations of SU(3) and SU(2), as in the Standard Model itself. On the other hand, a chiral SU(2) ordinary matter content could be still allowed, since SU(2) is anomaly free.

 Finally, we should briefly comment on  phenomenological properties of the one-loop, order $\theta$ renormalisable models found in this paper. There is ample literature regarding phenomenological studies of noncommutative theories in the enveloping algebra approach, but to our knowledge, all studies so far concerned models that are not GUT compatible. For a very limited list of phenomenological studies, we can point out those dealing with Standard Model forbidden decays, such as $Z\rightarrow\ga\ga$ \cite{Buric:2007qx}, $K\rightarrow\pi\ga$  \cite{Melic:2005su}, or quarkonia decays \cite{Tamarit:2008vy}, whose detection could be taken as a signal of noncommutativity. Other studies have dealt with noncommutative couplings between neutrinos and photons, such as in ref.~\cite{Schupp:2002up}; more collider phenomenology was analysed for example in ref.~\cite{Alboteanu:2006hh}. The phenomenological study of GUT compatible theories certainly deserves separate efforts, and we hope that our results will encourage them. In general, right-handed and left-handed fermions in GUT-compatible models couple to the noncommutativity parameters in different ways \cite{Martin:2009sg,Aschieri:2002mc}, in contrast to the case of theories constructed in terms of noncommutative Dirac fermions, which have been the focus of phenomenological studies so far. Thus, the results available in the literature cannot be directly extrapolated to the GUT-compatible case. Here we will just point out that in the renormalisable GUT-inspired models found in this paper and in ref.~\cite{Martin:2009vg},  there are no triple gauge interactions mixing gauge bosons belonging to different gauge groups, and also the couplings considered between matter and fermions in eq.~\eqref{S} can not mix neutrinos and photons. Hence, in an embedding of the Standard Model  into theories similar to those in this paper and in ref.~\cite{Martin:2009vg}, there would be no tree-level induced $Z\rightarrow\ga\ga$ decays, and no effects coming from tree-level neutrino-photon interactions. Standard Model forbidden decays involving kaons and quarkonia might still be allowed, but in the absence of $Z\rightarrow\ga\ga$ couplings the effects might be weaker, and there could be further cancellations due to the fact that the gauge interactions at first order in $\theta$ of left  handed fermions have opposite sign to those of right handed fermions.
	
\section{Acknowledgements}	 
	This work has been financially supported in part by MICINN through grant
FPA2008-04906, by both MICINN and the Fulbright Program through grant 2008-0800, and by the National Science Foundation under Grant No. PHY05-51164. The author wishes to thank C.~P.~Mart\'{\i}n for fruitful discussions and suggestions.
\newpage
	\appendix
\section{Feynman rules}
  We denote the background field legs with an encircled ``b"; the rules are defined without symmetrising over background field legs, in accordance with eq.~\eqref{Gammaexp}. Barred objects are strictly four-dimensional.
\subsection{U(1) theory}
The photon fields $a_\m^0$ --see eq.~\eqref{genexp}-- are denoted with a wavy line. The fermion multiplet $\psi_{i s}$ --``$i$" denoting the Dirac index and ``$s$" being the index of the representation-- are represented by single solid lines. \par
\psfrag{p}{$p$}
\psfrag{n B}{$\nu$}
\psfrag{r C}{$\r$}
\psfrag{m C}{$\m$}
\psfrag{i A}{$i s$}
\psfrag{j B}{$j t$}
\psfrag{q}{$q$}\psfrag{k}{$k$}
\psfrag{i s}{$i$}\psfrag{j t}{$j$}
\psfrag{k1}{$k_1$}\psfrag{k2}{$k_2$}\psfrag{k3}{$k_3$}
\begin{minipage}{0.27\textwidth}
\psfrag{m A}{\hskip10pt$\mu$}
\ \hskip-1cm\includegraphics[scale=1]{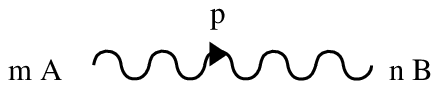}\hfill
\end{minipage}%
\begin{minipage}{0.23\textwidth}
\begin{align*}
\hskip-2cm\leftrightarrow\frac{-ig^{\m\n}}{p^2+i\epsilon}
\end{align*}
\end{minipage}%
\begin{minipage}{0.3\textwidth}
\includegraphics[scale=0.65]{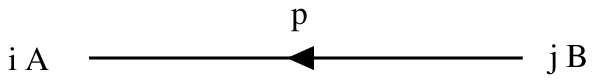}
\end{minipage}%
\begin{minipage}{0.16\textwidth}
\begin{align*}
\leftrightarrow\frac{i (\pslash)_{ij}\delta_{st}}{p^2+i\epsilon}
\end{align*}
\end{minipage}\\
\begin{minipage}{0.22\textwidth}
\flushleft
\includegraphics[scale=0.55]{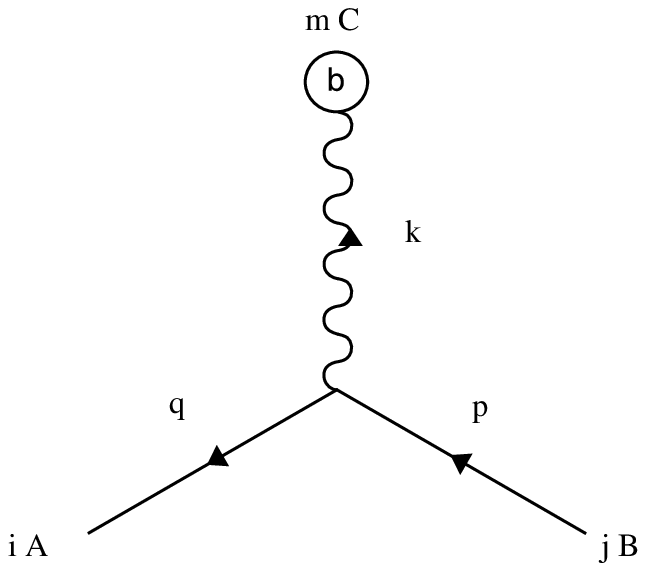}
\end{minipage}%
\begin{minipage}{0.2\textwidth}
\begin{align*}
&\leftrightarrow ie(\bar\ga^\m)_{ij}\rho_\psi(Y)_{st}
\end{align*}
\end{minipage}
\begin{minipage}{0.22\textwidth}
\flushleft
\includegraphics[scale=0.55]{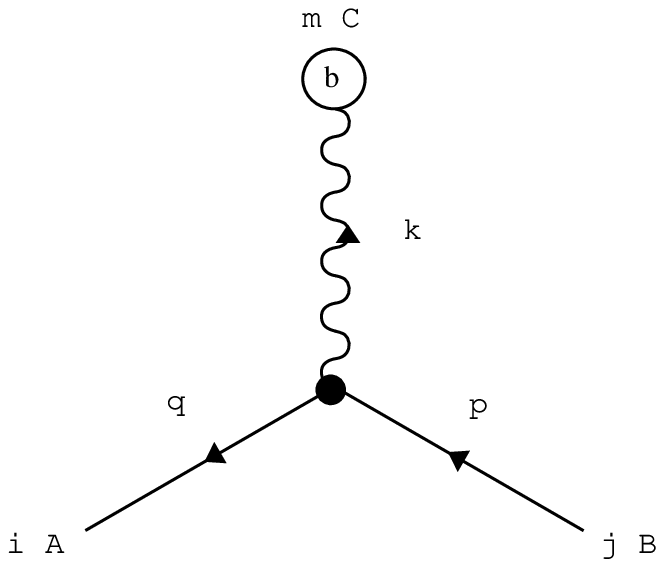}%
\end{minipage}%
\begin{minipage}{0.36\textwidth}%
\begin{align*}
\leftrightarrow\frac{e}{2}(\bg^\n P_L)_{ij}\th^{\a\b}\rho_\psi
(Y)_{st}[-(\bq_\n \bp_\b\\
-\bq_\b \bp_\n){\bdelta^\m}_\a)-\bk_\a \bp_\b{\bdelta^\m}_\n]
\end{align*}
\end{minipage}
\vskip0.4cm
\begin{minipage}{0.27\textwidth}\psfrag{m A}{\hskip10pt$\mu$}\psfrag{n B}{\hskip10pt$\n$}\hskip-1cm
\includegraphics[scale=0.6]{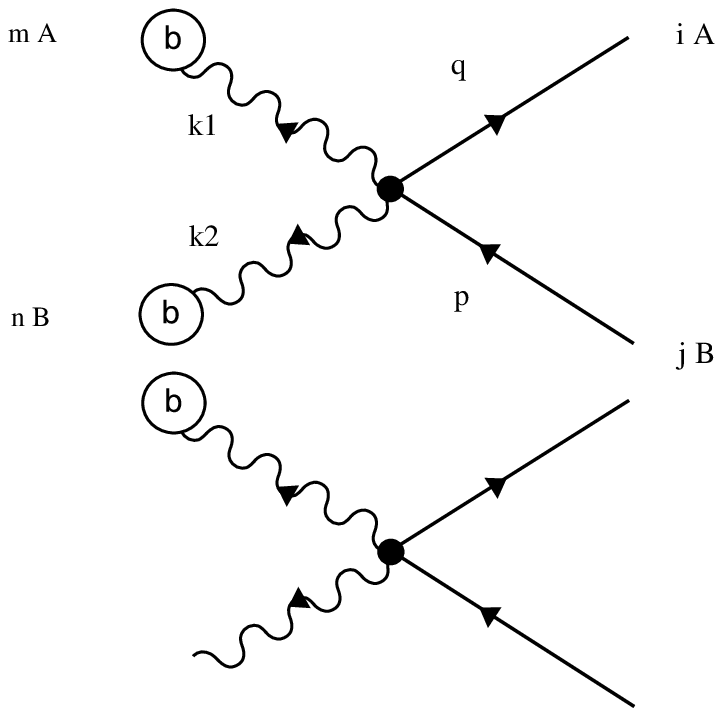}%
\end{minipage}%
\begin{minipage}{0.68\textwidth}%
\begin{align*}
\leftrightarrow&-\frac{e^2}{2}(\bg^\s P_L)_{ij}\th^{\a\b}(-(\bk_1-\bk_2)_\s\bdelta^{\m}_\a\bdelta^\n_\b\\
&+2\bk_{1\a}\bdelta^\m_\s\bdelta^\n_\b+2\bk_{2\a}\bdelta^\n_\s\bdelta^\m_\b-(\bq-\bp)_\b(\bdelta^\m_\a\bdelta^\n_\s+\bdelta^\n_\a\bdelta^\m_\s)).
\end{align*}
\end{minipage}

\newpage
\subsection{SU(N),  $\rm N>2$  theory}
The nonabelian gauge field $a_\m^a$ is denoted with a curly line, and he fermion multiplet $\psi_{is}$ is represented by double solid lines.\par
\psfrag{p}{$p$}
\psfrag{m A}{$\mu, a$}
\psfrag{n B}{$\nu, b$}
\psfrag{r C}{$\r, c$}
\psfrag{m C}{$\m, a$}
\psfrag{i A}{$i s$}
\psfrag{j B}{$j t$}
\psfrag{q}{$q$}\psfrag{k}{$k$}
\psfrag{i s}{$is$}\psfrag{j t}{$j t$}\psfrag{k,a,m}{$a,\mu$}
\psfrag{k1}{$k_1$}\psfrag{k2}{$k_2$}\psfrag{k3}{$k_3$}
\psfrag{m1}{$a,\mu$}\psfrag{m2}{\hskip0.2cm$b,\nu$}
\psfrag{l C}{$\la, c$}
\psfrag{r D}{$\r, d$}
\begin{minipage}{0.27\textwidth}
\hskip-.6cm
\includegraphics[scale=1]{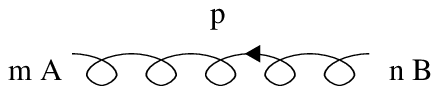}
\end{minipage}%
\begin{minipage}{0.23\textwidth}
\begin{align*}
\leftrightarrow\frac{-ig^2 \delta^{ab}g^{\m\n}}{p^2+i\epsilon}
\end{align*}
\end{minipage}%
\begin{minipage}{0.3\textwidth}
\includegraphics[scale=0.7]{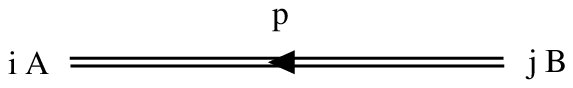}
\end{minipage}%
\begin{minipage}{0.16\textwidth}
\begin{align*}
\leftrightarrow\frac{i (\pslash)_{ij}\delta_{st}}{p^2+i\epsilon}
\end{align*}
\end{minipage}\\
\begin{minipage}{0.27\textwidth}
\flushleft\includegraphics[scale=0.6]{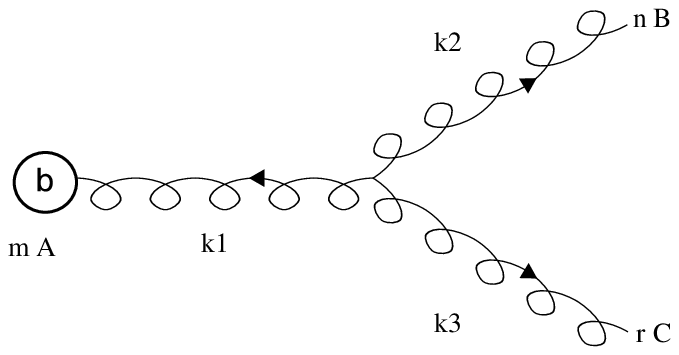}
\end{minipage}%
\begin{minipage}{0.33\textwidth}\flushleft
\begin{align*}
\leftrightarrow&\frac{1}{ g^2}f^{abc}[\bar g^{\m\r}(\bar k_1-\bar k_3-\bar k_2)^\n\\
&+\bar g^{\n\r}(\bar k_3-\bar k_2)^\m\\
&+\bar g^{\m\n}(\bar k_2-\bar k_1+\bar k_3)^\r]
\end{align*}
\end{minipage}
\begin{minipage}{0.22\textwidth}
\flushleft
\includegraphics[scale=0.55]{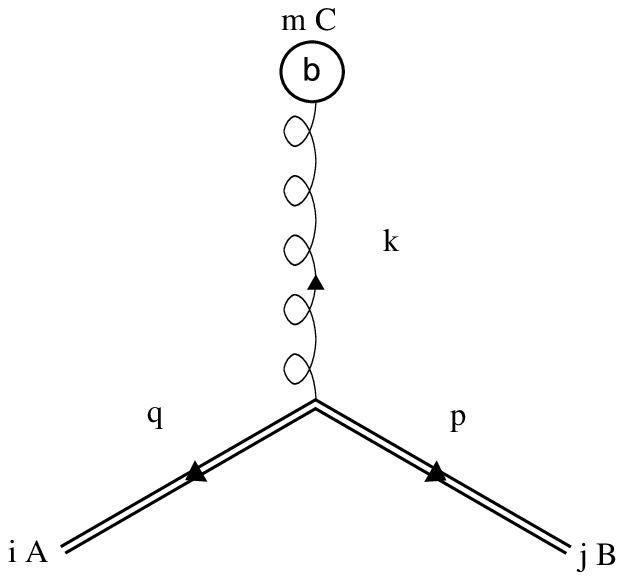}
\end{minipage}%
\begin{minipage}{0.18\textwidth}
\begin{align*}
&\leftrightarrow i(\bar\ga^\m)_{ij}(T^a)_{st}
\end{align*}
\end{minipage}\\
\begin{minipage}{0.4\textwidth}
\flushleft
\includegraphics[scale=0.7]{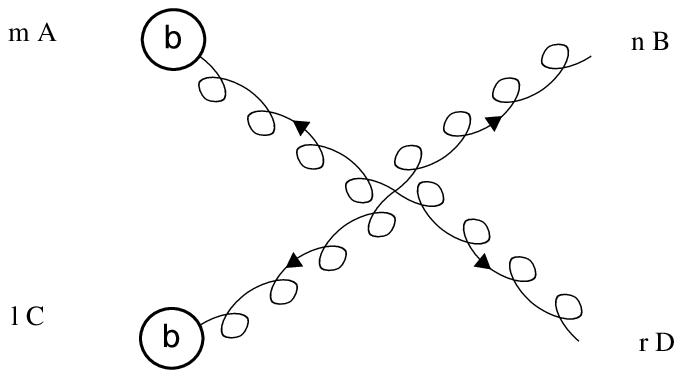}
\end{minipage}%
\begin{minipage}{0.56\textwidth}\flushleft
\begin{align*}
\leftrightarrow-\frac{i}{2 g^2}[&f^{abe}f^{ecd}(\bar g^{\m\la}\bar g^{\n\r}-\bar g^{\m\r}\bar g^{\n\la}+\bar g^{\m\n}\bar g^{\la\r})\\
&f^{ade}f^{ebc}(\bar g^{\m\n}\bar g^{\la\r}-\bar g^{\m\la}\bar g^{\n\r}-\bar g^{\m\r}\bar g^{\n\la})]\\
&f^{ace}f^{ebd}(\bar g^{\m\n}\bar g^{\la\r}-\bar g^{\m\r}\bar g^{\n\la})]
\end{align*}
\end{minipage}\\
\begin{minipage}{0.33\textwidth}
\flushleft\hskip0.1cm
\includegraphics[scale=0.7]{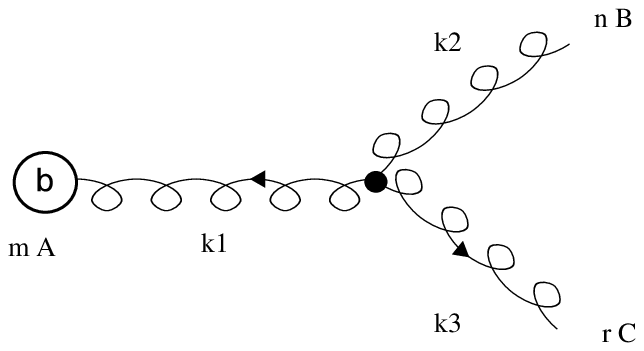}
\end{minipage}%
\begin{minipage}{0.63\textwidth}
\begin{align*}
\nonumber\leftrightarrow&\frac{1}{4\tilde g^2}\th_{\a\b}d^{abc}[\bar k_1^{\a}\bar k_2\cdot \bar k_3\bar g^{\m\b}\bar g^{\n\r}-\bar k_1^\a \bar k_2^\r \bar k_3^\n \bar g^{\m\b}\\
&-2(\bar k_1^\a \bar k_2^\b \bar k_3^\m \bar g^{\n\r}-\bar k_1^\a \bar k_2^\r \bar k_3^\m \bar g^{\n\b}-\bar k_1\cdot \bar k_3 \bar k_2^\b \bar g^{\n\r}\bar g^{\m\a}\\
\nonumber&+\bar k_1\cdot \bar k_3 \bar k_2^\r \bar g^{\m\a}\bar g^{\n\b})]+(\text{permutations of all legs})
\end{align*}
\end{minipage}\\
\begin{minipage}{0.4\textwidth}
\flushleft\hskip-0.2cm
\includegraphics[scale=0.7]{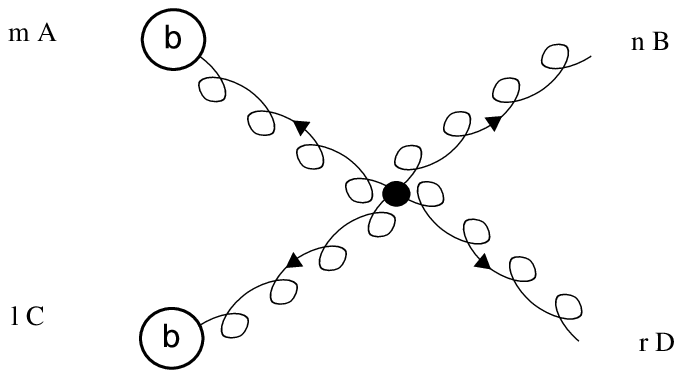}
\end{minipage}%
\begin{minipage}{0.56\textwidth}
\begin{align*}
\leftrightarrow&\frac{-i}{16\tilde g^2}\th_{\a\b}f^{abe}d^{cde}[\bar k_3\cdot \bar k_4 \bar g^{\m\a}\bar g^{\n\b}\bar g^{\la\r}-\bar k_3^\r \bar k_4^\la \bar g^{\m\a}\bar g^{\n\b}\\
&+4\bar k_3^\a \bar k_4^\m \bar g^{\n\r}\bar g^{\la\b}-4(\bar k_3^\b \bar k_4 ^\n \bar g^{\m\a}\bar g^{\la\r}-\bar k_3^\b \bar k_4 ^\la \bar g^{\m\a}\bar g^{\n\r}\\
&-\bar k_3^\r \bar k_4 ^\n \bar g^{\m\a}\bar g^{\la\b}+\bar k_3\cdot \bar k_4 \bar g^{\m\a}\bar g^{\n\r}\bar g^{\la\b})-2(\bar k_3^\a \bar k_4^\b \bar g^{\m\la}\bar g^{\n\r}\\
&-\bar k_3^\a \bar k_4^\n \bar g^{\m\la}\bar g^{\r\b}-\bar k_3^\m \bar k_4^\b \bar g^{\n\r}\bar g^{\la\a}+\bar k_3^\m \bar k_4^\n \bar g^{\la\a}\bar g^{\r\b})]\\
&+(\text{permutations of all legs})
\end{align*}
\end{minipage}\\\vskip0.1cm
\begin{minipage}{0.3\textwidth}
\flushleft
\includegraphics[scale=0.55]{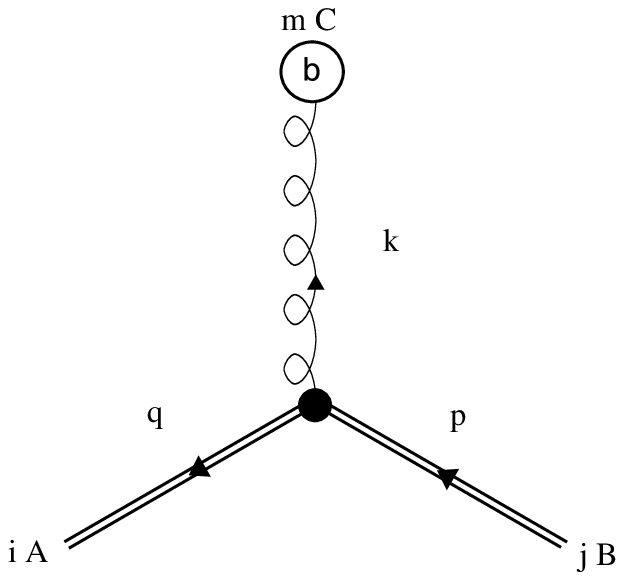}%
\end{minipage}%
\begin{minipage}{0.66\textwidth}%
\begin{align*}
\leftrightarrow\frac{1}{2}(\bg^\n P_L)_{ij}\th^{\a\b}\rho_\psi
(T^a)_{st}[-(\bq_\n \bp_\b-\bq_\b \bp_\n){\bdelta^\m}_\a)-\bk_\a \bp_\b{\bdelta^\m}_\n]
\end{align*}
\end{minipage}
\vskip0.4cm
\begin{minipage}{0.24\textwidth}\hskip-.8cm
\includegraphics[scale=0.6]{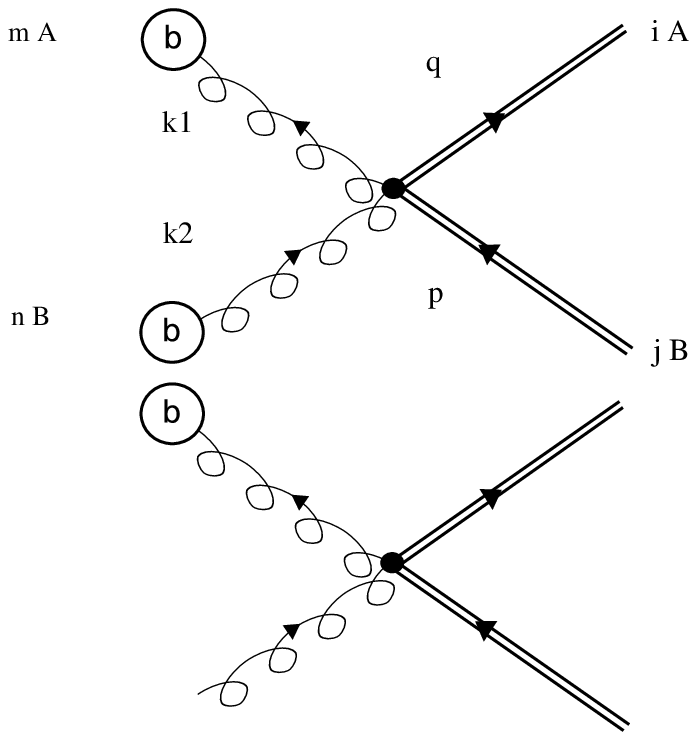}%
\end{minipage}%
\begin{minipage}{0.72\textwidth}%
\begin{align*}
\leftrightarrow&-\frac{1}{4}(\bg^\s P_L)_{ij}\th^{\a\b}\{\rho_\psi(T^a),\rho_\psi(T^b)\}_{st}(-(\bk_1-\bk_2)_\s\bdelta^{\m}_\a\bdelta^\n_\b\\
&+2\bk_{1\a}\bdelta^\m_\s\bdelta^\n_\b+2\bk_{2\a}\bdelta^\n_\s\bdelta^\m_\b-(\bq-\bp)_\b(\bdelta^\m_\a\bdelta^\n_\s+\bdelta^\n_\a\bdelta^\m_\s))\\
&+[\rho_L(T^i)^{(a)},\rho_L(T^j)^{(b)}]_{st}((\bq+\bp)_\s\bdelta^\m_\a\bdelta^\n_\b-(\bq+\bp)_\b(\bdelta^\m_\a\bdelta^\n_\s-\bdelta^\n_\a\bdelta^\m_\s))].
\end{align*}
\end{minipage}


\section{$\beta$ functions for the physical couplings, $e,g,\th$}

The $\beta$ functions for the couplings of the U(1) and SU(N) theories can be readily shown to be
\begin{align*}
 U(1):& \quad \beta_e=\frac{e^3}{12\pi^2},\\
      &  \quad \beta_\th=\frac{2e^2\th}{3\pi^2},\\
 SU(N):& \quad \beta_g=-\frac{g^3}{16\pi^2}\Big(\frac{11}{3}N-\frac{4}{3}\sum_r c_2(r)\Big),\\
      &  \quad \beta_\th=-\frac{g^2\th}{6\pi^2}(N-4 C_2(r)).\\
\end{align*}
In the formulae above, $C_2(r)$ represents the second-degree Casimir invariant of the representation $r$, and $c_2(r)$ is the index of the representation. They are related by the relation
\begin{align*}
C_2(r)=c_2(r)\frac{N(G)}{N(r)},
\end{align*}
where $N(r)$ is the dimension of the representation $r$, and $G$ denotes the adjoint representation.
The $\beta$ functions for the gauge couplings $e,g$ are the same as in the commutative theory. The $\beta$ functions for $\theta$ are generically positive in the presence of non-singlet matter; in the SU(3) case, for example, using the tables in ref.~\cite{Slansky:1981yr}, $\beta_\th$ is positive for representations of dimension less than 65, and the same result applies for SU(5) representations of dimension less than $800$.

\end{document}